\documentclass[longauth]{aa}  
\usepackage{graphicx}
\usepackage{txfonts}
\usepackage{caption}
\usepackage{placeins}
\usepackage{amsmath}	
\usepackage{amssymb}
\usepackage{threeparttablex}
\usepackage{hyperref}
\hypersetup{
  colorlinks   = true, 
  urlcolor     = blue, 
  linkcolor    = blue, 
  citecolor   = blue 
}
\usepackage{natbib}
\usepackage{orcidlink}
\bibpunct{(}{)}{;}{a}{}{,}
\begin{document}

   \title{TOI-5108\,b and TOI 5786\,b: Two transiting sub-Saturns detected and characterized with TESS, MaHPS and SOPHIE}
   
   \author{Luis Thomas\thanks{Email:lthomas@mpe.mpg.de}\orcidlink{0009-0006-1571-0306}\inst{1,2}
          \and
          Guillaume H\'ebrard \inst{3,4}
          \and 
          Hanna Kellermann \orcidlink{0009-0006-3527-0424} \inst{1,2}
          \and
          Judith Korth \orcidlink{0000-0002-0076-6239} \inst{5}
          \and
          Neda Heidari \orcidlink{0000-0002-2370-0187} \inst{3} 
          \and
          Thierry Forveille \orcidlink{0000-0003-0536-4607} \inst{6}
          \and
          S\'ergio G. Sousa \inst{7}
          \and
          Laura Schöller \orcidlink{0009-0008-2469-0372} \inst{1}
          \and
          Arno Riffeser \inst{1}
          \and
          Claus Gössl \inst{1}
          \and
          Juan Serrano Bell \orcidlink{0000-0002-8397-557X} \inst{8}
          \and
          Flavien Kiefer \inst{9}
          \and
          Nathan Hara \inst{10}
          \and
          Frank Grupp \inst{1,2}
          \and
          Juliana Ehrhardt \orcidlink{0009-0003-9433-043X} \inst{1,2}
          \and
          Felipe Murgas \orcidlink{0000-0001-9087-1245} \inst{11,12}
          \and
          Karen A. Collins \orcidlink{0000-0001-6588-9574} \inst{13}
          \and
          Allyson Bieryla \orcidlink{0000-0001-6637-5401} \inst{13}
          \and
          Hannu Parviainen \orcidlink{0000-0001-5519-1391} \inst{11,12}
          \and
          Alexandr A. Belinski \orcidlink{0000-0003-3469-0989} \inst{14}
          \and
          Emma Esparza-Borges \orcidlink{0000-0002-2341-3233} \inst{11,12}
          \and
          David R. Ciardi \orcidlink{0000-0002-5741-3047} \inst{15}
          \and
          Catherine A. Clark \orcidlink{0000-0002-2361-5812} \inst{15}
          \and
          Akihiko Fukui \orcidlink{0000-0002-4909-5763} \inst{16,11}
          \and
          Emily A. Gilbert \orcidlink{0000-0002-0388-8004} \inst{17}
          \and
          Ulrich Hopp \inst{1}
          \and
          Kai Ikuta \orcidlink{0000-0002-5978-057X} \inst{18}
          \and
          Jon M. Jenkins \orcidlink{0000-0002-4715-9460} \inst{19}
          \and
          David W. Latham \orcidlink{0000-0001-9911-7388} \inst{13}
          \and
          Norio Narita \orcidlink{0000-0001-8511-2981} \inst{16,20,11}
          \and
          Louise D. Nielsen \orcidlink{0000-0002-5254-2499} \inst{1}
          \and
          Samuel N. Quinn \orcidlink{0000-0002-8964-8377} \inst{13}
          \and
          Enric Palle \orcidlink{0000-0003-0987-1593} \inst{11,12}
          \and
          Jan-Niklas Pippert \orcidlink{0009-0006-9461-002X} \inst{1,2}
          \and
          Alex S. Polanski \orcidlink{0000-0001-7047-8681} \inst{21,22}
          \and
          Christoph Ries \inst{1}
          \and
          Michael Schmidt \inst{1}
          \and
          Richard P. Schwarz \orcidlink{0000-0001-8227-1020} \inst{13}
          \and
          Sara Seager \orcidlink{0000-0002-6892-6948}  \inst{23,24,25}
          \and
          Ivan A. Strakhov \orcidlink{0000-0003-0647-6133} \inst{14}
          \and
          Stephanie Striegel \orcidlink{0009-0008-5145-0446} \inst{26}
          \and
          Julian C. van Eyken \orcidlink{0000-0003-2192-5371} \inst{15}
          \and
          Noriharu Watanabe \orcidlink{0000-0002-7522-8195} \inst{18}
          \and
          Cristilyn N. Watkins \orcidlink{0000-0001-8621-6731} \inst{13}
          \and
          Joshua N. Winn \orcidlink{0000-0002-4265-047X} \inst{27}
          \and
          Carl Ziegler \orcidlink{0000-0002-0619-7639} \inst{28}
          \and
          Raphael Zöller \orcidlink{0000-0002-0938-5686} \inst{1,2}
}

   \institute{
   University Observatory Munich, Faculty of Physics, Ludwig-Maximilians-Universit\"at München, Scheinerstr. 1, 81679 Munich, Germany
    \and
    Max-Planck Institute for Extraterrestrial Physics, Giessenbachstrasse 1, D-85748 Garching, Germany
    \and
    Institut d'astrophysique de Paris, UMR7095 CNRS, Universit\'e Pierre \& Marie Curie, 98bis boulevard Arago, 75014 Paris, France
    \and
    Observatoire de Haute-Provence, CNRS, Universit\'e d'Aix-Marseille, 04870 Saint-Michel-l'Observatoire, France   
    \and
    Lund Observatory, Division of Astrophysics, Department of Physics, Lund University, Box 118, 22100 Lund, Sweden
    \and
    Universit\'e Grenoble Alpes, CNRS, IPAG, 38000 Grenoble, France
    \and
    Instituto de Astrof{\'\i}sica e Ci\^encias do Espa\c{c}o, Universidade do Porto, CAUP, Rua das Estrelas, 4150-762 Porto, Portugal
    \and
    International Center for Advanced Studies (ICAS) and ICIFI (CONICET), ECyT-UNSAM, Campus Miguelete, 25 de Mayo y Francia, (1650) Buenos Aires, Argentina
    \and
    LESIA, Observatoire de Paris, Universit\'e PSL, CNRS, Sorbonne Universit\'e, Universit\'e Paris Cit\'e, 5 place Jules Janssen, 92195 Meudon, France
    \and
    Aix Marseille Universit\'e, CNRS, LAM (Laboratoire d'Astrophysique de Marseille) UMR 7326, 13388 Marseille, France
    \and 
    Instituto de Astrof\'{i}sica de Canarias (IAC), 38205 La Laguna, Tenerife, Spain 
    \and 
    Departamento de Astrofísica, Universidad de La Laguna (ULL), 38206, La Laguna, Tenerife, Spain 
    \and
    Center for Astrophysics \textbar \ Harvard \& Smithsonian, 60 Garden Street, Cambridge, MA 02138, USA
    \and
    Sternberg Astronomical Institute, M.V. Lomonosov Moscow State University, 13, Universitetskij pr., 119234, Moscow, Russia
    \and
    NASA Exoplanet Science Institute-Caltech/IPAC, Pasadena, CA 91125, USA
    \and
    Komaba Institute for Science, The University of Tokyo, 3-8-1 Komaba, Meguro, Tokyo 153-8902, Japan
    \and
    Jet Propulsion Laboratory, California Institute of Technology, 4800 Oak Grove Drive, Pasadena, CA 91109, USA
    \and
    Department of Multi-Disciplinary Sciences, Graduate School of Arts and Sciences, The University of Tokyo, 3-8-1 Komaba, Meguro, Tokyo 153-8902, Japan
    \and
    NASA Ames Research Center, Moffett Field, CA 94035 USA
    \and
    Astrobiology Center, 2-21-1 Osawa, Mitaka, Tokyo 181-8588, Japan
    \and
    Lowell Observatory, 1400 W Mars Hill Road, Flagstaff, AZ, 86001, USA
    \and
    Department of Physics and Astronomy, University of Kansas, Lawrence, KS 66045, USA
    \and
    Department of Earth, Atmospheric, and Planetary Sciences, Massachusetts Institute of Technology, Cambridge, MA 02139, USA
    \and
    Department of Physics and Kavli Institute for Astrophysics and Space Research, Massachusetts Institute of Technology, Cambridge, MA 02139, USA
    \and
    Department of Aeronautics and Astronautics, Massachusetts Institute of Technology, Cambridge, MA 02139, USA
    \and
    SETI Institute, Mountain View, CA 94043 USA
    \and
    Department of Astrophysical Sciences, Princeton University, Princeton, NJ 08544, USA
    \and
    Department of Physics, Engineering and Astronomy, Stephen F. Austin State University, 1936 North St, Nacogdoches, TX 75962, USA
    }
   \date{}

 \titlerunning{TOI-5108\,b and TOI 5786\,b: Two sub-Saturns detected and characterized with TESS, MaHPS and SOPHIE}
  \abstract
  {We report the discovery and characterization of two sub-Saturns from the Transiting Exoplanet Survey Satellite (\textit{TESS}) using high-resolution spectroscopic observations from the MaHPS spectrograph at the Wendelstein Observatory and the SOPHIE spectrograph at the Haute-Provence Observatory. Combining photometry from TESS, KeplerCam, LCOGT, and MuSCAT2 with the radial velocity measurements from MaHPS and SOPHIE we measure precise radii and masses for both planets. TOI-5108\,b is a sub-Saturn with a radius of $6.6 \pm 0.1$ $R_\oplus$ and a mass of $32 \pm 5$ $M_\oplus$. TOI-5786\,b is similar to Saturn with a radius of $8.54 \pm 0.13$ $R_\oplus$ and a mass of $73 \pm 9$ $M_\oplus$. The host star for TOI-5108~b is a moderately bright (Vmag 9.75) G-type star. TOI-5786 is a slightly dimmer (Vmag 10.2) F-type star. Both planets are close to their host stars with periods of 6.75~days and 12.78~days respectively. This puts TOI-5108\,b just inside the bounds of the Neptune desert while TOI-5786\,b is right above the upper edge. We estimate hydrogen-helium envelope mass fractions of $38\,\%$ for TOI-5108\,b and $74\,\% $ for TOI-5786\,b. However, using a model for the interior structure that includes tidal effects the envelope fraction of TOI-5108\,b could be much lower ($\sim 20\,\%$) depending on the obliquity. We estimate mass-loss rates between 1.0 $\times 10^9$\,g/s and 9.8 $\times 10^9$\,g/s for TOI-5108\,b and between $3.6 \times 10^8$\,g/s and $3.5 \times 10^9$\,g/s for TOI-5786\,b. Given their masses, this means that both planets are stable against photoevaporation. Furthermore, at these mass-loss rates, there is likely no detectable signal in the metastable helium triplet with the James Webb Space Telescope. We also detect a transit signal for a second planet candidate in the TESS data of TOI-5786 with a period of 6.998 days and a radius of $3.83 \pm 0.16$ $R_\oplus$. Using our RV data and photodynamical modeling, we are able to provide a 3-$\sigma$ upper limit of 26.5 $M_\oplus$ for the mass of the potential inner companion to TOI-5786\,b.}






   \maketitle
%

\section{Introduction}

A remarkable diversity has been found among the over 5700 exoplanets that are known thus far. Interestingly, the most common exoplanets in the sample of known exoplanets, sub-Neptunes with radii between $2-4 \ R_\oplus$, have no counterpart in our solar system. Another planet type that is found in the exoplanet population and not present in the Solar system are planets with sizes between Neptune and Saturn ($4$ $R_\oplus < R_p < 9$ $R_\oplus$), often called super-Neptunes, sub-Saturns or sub-Jovian planets (hereinafter sub-Saturns). Compared to the smaller planets like super-Earths and sub-Neptunes, sub-Saturns are much rarer with the occurrence rate of planets dropping sharply for planet sizes above $\sim 3$ $R_\oplus$ \citep{Fulton2018,Hsu2019,Kite2019}. The formation and evolution of sub-Saturns is therefore expected to differ from the more common sub-Neptunes. 
\\
While sub-Saturns are expected to have a substantial H/He envelope they seem to have remarkably little correlation between their size and masses spanning a wide range of bulk densities \citep{Petigura2017,Nowak2020,Mistry2023}. The interior structure of these planets can be well described with a two-component model including an Earth-composition core and a H/He envelope. Their size is then largely determined by the envelope fraction $f_{env} = M_{env} / M_{tot}$ with changes in the detailed composition of the core only having minor effects on the size of the planet \citep{Lopez2014,Petigura2016}.
\\
Using models such as \cite{Lopez2014}, which considers various sources of heating and cooling of the envelope over the planet's lifetime given the incident stellar flux, the observed masses and radii of the planets can be converted to estimates of the envelope fraction. For some of the low-density sub-Saturns this results in envelope fractions $ \sim 50 \%$ or higher \citep[e.g. ][]{Petigura2018a,Petigura2020,Mistry2023}. 
\\
The larger envelope fractions conflict with the prevailing theory for the formation of sub-Saturns which suggests that these planets do not reach the point of runaway gas accretion explaining their smaller size compared to Jovian planets. In that sense, they could be considered failed gas giants. As runaway accretion is expected to start when $M_{core} \approx M_{env}$ ($f_{env} \approx 50 \%$), sub-Saturns with such envelope fractions should have triggered runaway accretion and grown to much larger sizes very quickly. While it is possible that these planets reached their present-day envelope fractions just as the protoplanetary disk dispersed, the required fine-tuning in the timing would mean that such planets should be rare. \cite{Millholland2020} proposed tidal heating as a mechanism to inflate the radii of these sub-Saturns which can lead the models to overestimate the envelope fractions from the observed radii. They found that the envelope fractions of sub-Saturns are on average $\sim 10 \%$ smaller when tidal heating is taken into account and nearly half of the population would have radii consistent with sub-Neptunes if not for tidal inflation. 
\\
An alternative explanation proposed by \cite{Helled2023} is that runaway accretion is delayed by an intermediate stage of efficient heavy element accretion and only starts at a planetary mass $\sim 100$ $M_\oplus$ meaning that even Saturn would not have started runaway accretion. This can explain the higher bulk metallicity of Saturn compared to Jupiter which has also been found in some of the Saturn-like exoplanets \citep{Han2024}.
\\
Another interesting feature that is seen in the population of intermediate-sized planets is the relative scarcity of sub-Saturns in very close orbits ($P<5$ days) called the Neptune desert \citep{lecavelier2007,szabo2011,nesvorn2013,ojeda2014,lundkvist2016,Mazeh2016}. 
While different atmospheric escape mechanisms such as photoevaporation or Roche Lobe Overflow have been proposed as the explanation of the Neptune desert \citep{kurokawa2014,jackson2017,owen2019,koskinen2022}, there are indications that two separate mechanisms are responsible for the upper and lower bound of the Neptune desert \citep{ionov2018,owenlai2018}. At the lower boundary photoevaporation is efficient in stripping Neptune-sized planets of their atmospheres, leaving them as bare cores with little to no atmosphere \citep{lecavelier2007,owenwu2013}.
However, the planets at the upper edge seem to be stable against photoevaporation due to their deeper gravitational wells \citep{viss2022}. \cite{owenlai2018} suggested that this part of the desert is sculpted by constraints of high-eccentricity migration where tidal forces from the star limit the masses of planets that can survive in very close orbits \citep[see also][]{koenig2016}. 
\\
Insights into the underlying mechanisms shaping the Neptune desert can come from precisely determining the planetary, stellar, and orbital parameters for planets around and inside the desert. Such observations suggest that planets around the lower edge of the Neptune desert are more common around higher metallicity host stars which would be expected if photoevaporation is the driving process. This is consistent with higher metallicity exospheres undergoing less efficient photoevaporation due to enhanced cooling from atomic metal lines in the planetary atmosphere \citep{dong2018,petigura2018b,owenmurray2019}.
\\
If the upper edge is indeed caused by the effects of high-eccentricity migration it is expected that planets there show signs of long-period massive companions which can induce migration but lack nearby planetary companions as they would have been disturbed during the migration process. Furthermore, it would be expected to find a population of sub-Saturns at slightly larger periods with moderately to high eccentric orbits that are in the process of migrating inwards and circularizing \citep{correia2020}. In this case, the loss of the atmosphere through photoevaporation would be delayed so even Neptune-sized planets would be able to retain their atmosphere until they migrate further in. Two potential members of this class of late Neptunian migrators are  GJ\,436\,b and GJ\,3470\,b whose moderate eccentricity, measured evaporation, and misaligned orbits point to late-stage
Kozai–Lidov migration \citep{bourrier20182,bourrier20181,dossantos2019,palle2020,ninan2020,attia2021,stefansson2022}.
\\
Additional constraints on the origin of the Neptune desert can come from atmospheric characterization of planets in and around the desert. Tracers such as the Lyman-$\alpha$ line of hydrogen \citep{vidal2003,Lecavelier2010,Ehrenreich2015,Bourrier2017} or the helium triplet at 1083\,nm \citep{Oklop2018,Mansfield2018,Spake2018} can be used to measure atmospheric mass-loss rates in planets. These measurements can further our understanding at which points in the parameter space photoevaporation is efficient enough to have a significant impact on the evolution of a young planet. 
\\
In this work, we report the detection and characterization of two sub-Saturns TOI-5108\,b and TOI-5786\,b. We derive precise orbital and planetary parameters by combining photometric measurements from the Transiting Exoplanet Survey Satellite \citep[\textit{TESS},][]{Ricker2015} and ground-based observations with radial velocity observations from the SOPHIE and MaHPS high-resolution spectrographs. Furthermore, we report the signal of an additional transiting planet candidate in the \textit{TESS} lightcurve of TOI-5786. Section \ref{secobs} describes the observations conducted in this work. In Section \ref{stellar} we derive the parameters of the host stars from the photometric and spectroscopic data. In Section \ref{Analysis} we present the analysis of the planet properties for the three planet candidates. Finally, the derived planet properties are discussed in Section \ref{results_disc}.

\section{Observations} \label{secobs}

\subsection{TESS photometry}
\begin{figure*}
    \centering
    \includegraphics[width=2\columnwidth]{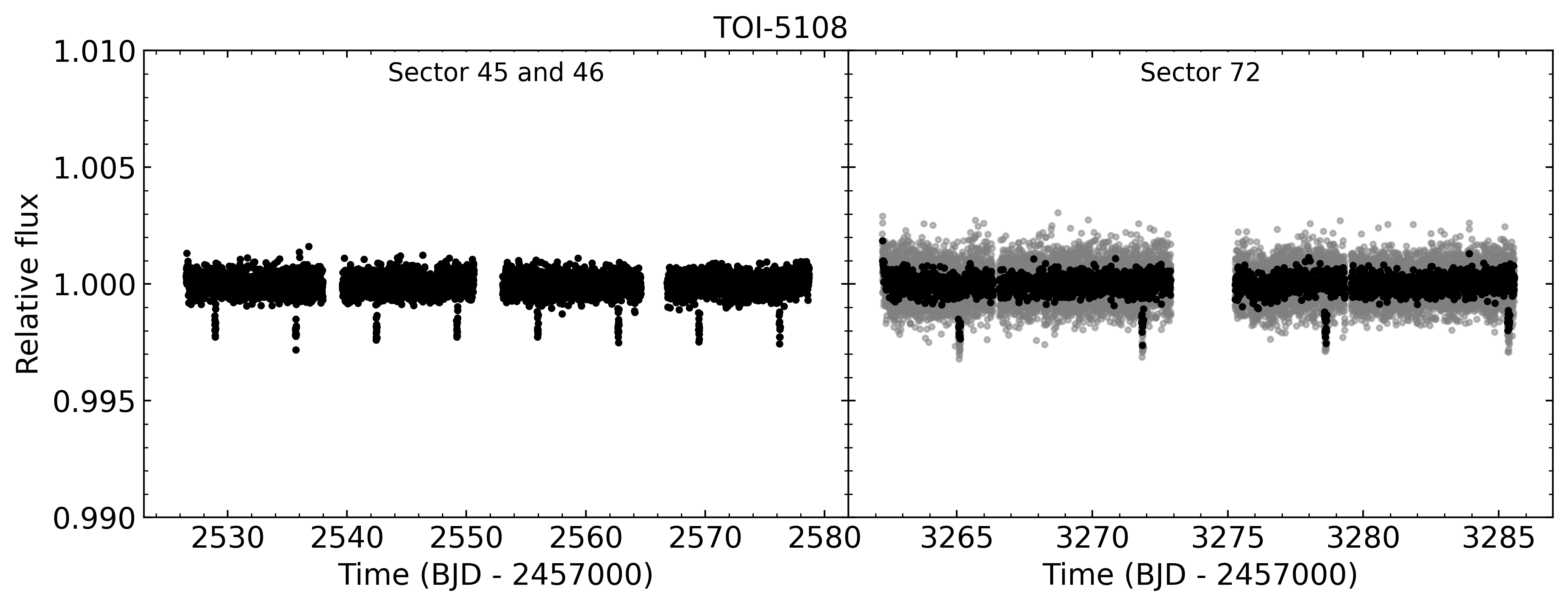} 
    \includegraphics[width=2\columnwidth]{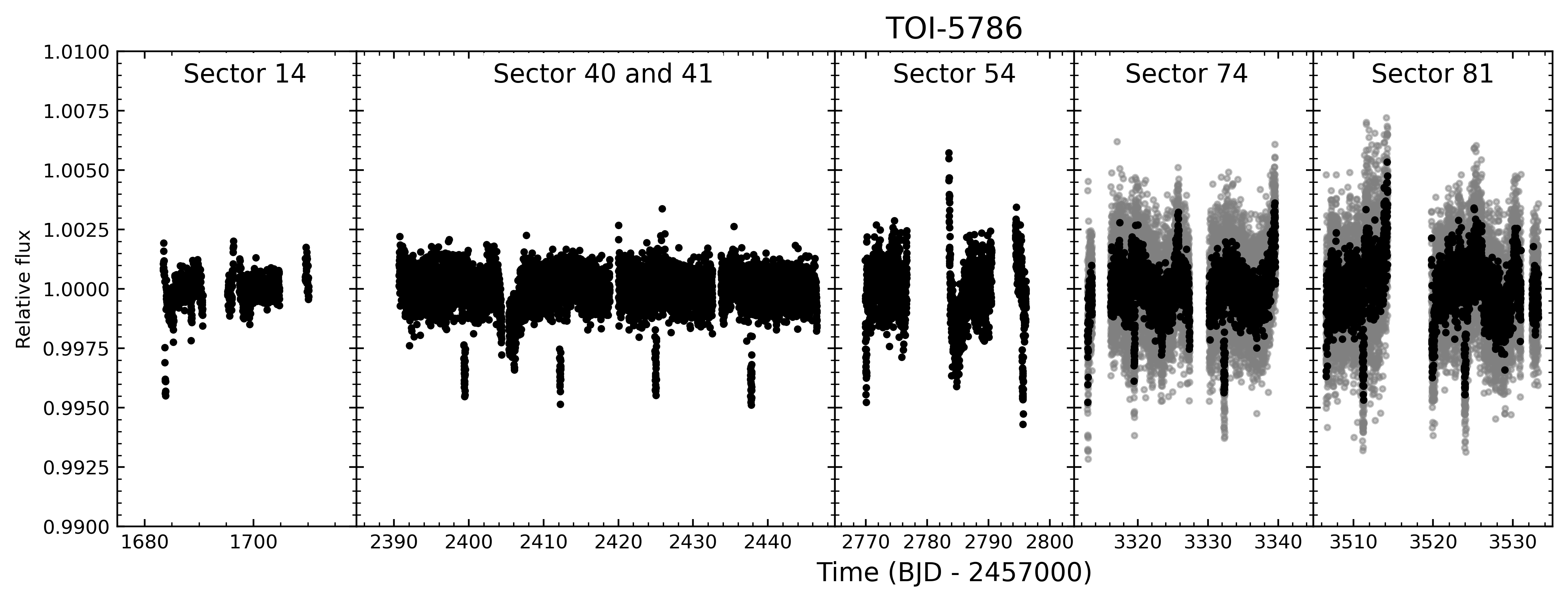} 
    \caption{PDCSAP from the \textit{TESS} lightcurves for TOI-5108 (upper plot) and TOI-5786 (lower plot). For the last sectors for both targets where high cadence photometry is available, we show the 120-second cadence in grey and the flux data binned to the 10-minute cadence in black.}
    \label{fig:tesslc}
\end{figure*}

\textit{TESS} observed TOI-5108 during the first extended mission when the spacecraft pointed at the ecliptic for the first time in Sector 45 from UT November 6 to December 2, 2021, and in Sector 46 from UT December 2 to December 30, 2021, with a 10-minute cadence. \textit{TESS} returned to the ecliptic during the second extended mission and observed TOI-5108 again in Sector 72 from UT November 11 to December 7, 2023, with a shorter 20-second cadence.
\\
TOI-5786 was observed in Sector 14 from UT July 18 to August 14, 2019, (30-minute cadence), Sector 40 from UT June 24 to July 23, 2021, Sector 41 from UT July 23 to August 20, 2021, Sector 54 from UT July 9 to August 5, 2022, (all at 10-minute cadence) as well as Sector 74 from UT January 3 to January 30, 2024, and Sector 81 from UT July 15 to August 10, 2024, with a 20-second cadence.
\\
For both stars, transit events were initially identified in the full frame images by the \textit{TESS} Quick Look Pipeline \citep[QLP, ][]{huang2020photometry,huang2020b,kunimoto2021,Kunimoto2022}. The \textit{TESS} Science Processing Operation Center pipeline \citep[SPOC][]{jenkins2016tess} at NASA Ames Research Center subsequently detected TOI-5108.01 and TOI-5786.01 after conducting a transit search of Sector 72 and Sector 40 with an adaptive, noise-compensating matched filter \citep{Jenkins2002,Jenkins2010,Jenkins2020}, producing a TCE for which an initial limb-darkened transit model was fitted \citep{Li2019}. Following that detection, light curves were produced as TESS-SPOC High-Level Science Product \citep{caldwell2020tess} and uploaded to the Mikulski Archive for Space Telescopes (MAST). After a series of diagnostic tests, both stars were released to the public as \textit{TESS} Objects of Interest (TOI).
\\
TOI-5108.01 is reported to have a period of $6.75357 \pm 0.00003$ days and a transit depth of $1740 \pm 60$ ppm. TOI-5786.01 has a period of $12.77919 \pm 0.00007$ with a depth of $3008 \pm 152$ ppm.
\\
For our analysis we used the \texttt{lightkurve} package \citep{2018ascl.soft12013L} to download the PDCSAP \citep[Presearch Data Conditioning Simple Aperture Photometry,][]{smith2012kepler,stumpe2012kepler,stumpe2014multiscale} light curves, which correct the Simple Aperture Photometry (SAP) data for instrumental systematics and also corrects for the contamination of known nearby stars. For the short cadence sectors, we use the 120-second lightcurves. The \textit{TESS} lightcurves for both targets are shown in Figure \ref{fig:tesslc}.

\subsection{Ground-based photometric follow-up}

The \textit{TESS} pixel scale is $\sim 21\arcsec$ per pixel and photometric apertures typically extend out to roughly 1\arcmin, generally causing multiple stars to blend in the \textit{TESS} photometric aperture. To attempt to determine the true source of the \textit{TESS} detections, we acquired ground-based time-series follow-up photometry of the field around TOI-5108 and TOI-5786 as part of the \textit{TESS} Follow-up Observing Program \citep[TFOP;][]{collins:2019}\footnote{https://tess.mit.edu/followup}. We used the {\tt TESS Transit Finder}, which is a customized version of the {\tt Tapir} software package \citep{Jensen:2013}, to schedule our transit observations. All light curve data are available on the {\tt EXOFOP-TESS} website\footnote{\href{https://exofop.ipac.caltech.edu/tess/target.php?id=350348197}{https://exofop.ipac.caltech.edu/tess/target.php?id=350348197}}. The lightcurves are shown in Figure \ref{fig:lcall}. A summary of the observations is given in Table \ref{tab:ground}.

\subsubsection{KeplerCam}
TOI-5108.01 was observed on UT April 1, 2022, with KeplerCam on the 1.2m telescope at the Fred Lawrence Whipple Observatory (FLWO) at Mount Hopkins, Arizona. KeplerCam is a $4096 \times 4096$ Fairchild CCD 486 detected which has a field-of-view of $23\arcmin\times23\arcmin$ and an image scale of 0.672\arcsec per pixel when binned by 2. Data were reduced using standard IDL routines and aperture photometry was performed using AstroImageJ \citep{collins:2017}. 

\subsubsection{LCOGT\label{subsubsec:lcogt}}
We observed a full transit window of TOI-5108.01 in Pan-STARRS $z$-short band on UT December 20, 2023, from the Las Cumbres Observatory Global Telescope (LCOGT) \citep{Brown:2013} 1\,m network node at McDonald Observatory near Fort Davis, Texas, United States (McD). The 1\,m telescope is equipped with a $4096\times4096$ SINISTRO camera having an image scale of $0.389\arcsec$ per pixel, resulting in a $26\arcmin\times26\arcmin$ field of view. The images were calibrated by the standard LCOGT {\tt BANZAI} pipeline \citep{McCully:2018}, and differential photometric data were extracted using {\tt AstroImageJ}.

\subsubsection{MuSCAT2}
We observed TOI-5108 and TOI-5786 with the multi-band imager MuSCAT2 \citep{Narita2019} mounted on the 1.5 m Telescopio Carlos S\'{a}nchez (TCS) at Teide Observatory, Spain. MuSCAT2 is equipped with four CCDs and can obtain simultaneous images in $g'$, $r'$, $i'$, and $z_s$ bands with little readout time. Each CCD has $1024 \times 1024$ pixels with a field of view of $7.4\arcmin \times 7.4\arcmin$. All the MuSCAT2 data were reduced by the MuSCAT2 pipeline \citep{Parviainen2019}. The pipeline performs standard calibrations (dark and flat field corrections), aperture photometry, and transit model fit including instrumental systematics.
\\
TOI-5786 was observed on UT July 14, 2023, and UT November 19, 2023. The TOI-5786 observations on the night of 14 July 2023 were made with nominal focus, and the exposure times were set to 5\,s in all bands. The observations cover the full transit event, although with very little pre- and post-transit data.
\\
The TOI-5786 observations taken on 19 November 2023 were made with the telescope defocused. The exposure times were set to 3\,s in $g'$, and 5\,s in $r'$, $i'$, and $z_s$, respectively. The full transit could not be observed, and the data cover the ingress with little pre-transit baseline.
\\
TOI-5108 was observed with MuSCAT2 on the night of 26 December 2023. The observations were made with the telescope defocused, and the exposure times were set to 15 seconds for all bands. The egress of the transit was detected.
\begin{table}
    \centering
    \caption{\label{tab:ground}Summary of the ground-based follow-up observations for TOI-5108 and TOI-5786.}
    \resizebox{\columnwidth}{!}{
    \begin{tabular}{lcccc}
    \hline
    \hline
      Date   & Instrument & Mirror [m]& Filter & Aperture [\arcsec]\\
      \hline
      \\
      TOI-5108  \\
      \\
      2022-04-01&KeplerCam& 1.2  & zp& 8.1 \\
      2023-12-20& SINISTRO&1.0 & zs& 7.0\\
      2024-12-26& MuSCAT2 & 1.52 & g,r,i,zs &10.875\\
      \\
      TOI-5786 \\
      \\
       2023-07-14& MuSCAT2 &1.52 & g,r,i,zs &10.875\\
        2024-11-19& MuSCAT2 & 1.52 & g,r,i,zs &10.875\\
    \hline
    \end{tabular}
    }
\end{table}
\subsection{High-angular-resolution imaging}
We also collected high-angular-resolution imaging observations for both of our targets to search for unidentified close companions. These companions could dilute the transit signal leading to an underestimated radius for the planet \citep{ciardi2015} or in the worst case create false-positive scenarios if they happen to be eclipsing binaries. 
\\
We searched for stellar companions to TOI-5108 with speckle imaging on the 4.1-m Southern Astrophysical Research (SOAR) telescope \citep{Tokovinin2018} on 15 April 2022 UT, observing in Cousins I-band, a similar visible bandpass as TESS. This observation was sensitive to a 5.3-magnitude fainter star at an angular distance of 1\arcsec from the target. More details of the observations within the SOAR \textit{TESS} survey are available in \cite{Ziegler2020}. The 5$\sigma$ detection sensitivity and speckle auto-correlation functions from the observations are shown in Figure \ref{fig:dirim}. No nearby stars were detected within 3\arcsec\ of TOI-5108 in the SOAR observations.
\\
We observed TOI-5786 on November 8, 2022, UT and again on December 30, 2023, UT with the speckle polarimeter on the 2.5-m telescope at the Caucasian Observatory of Sternberg Astronomical Institute (SAI) of Lomonosov Moscow State University. The speckle polarimeter uses a high--speed, low--noise Hamamatsu ORCA--quest CMOS detector \citep{Strakhov2023}. 
Both observations were carried out in the $I_\mathrm{c}$ band with an angular resolution of 0.083\arcsec, and seeing conditions of 0.6\arcsec  and 0.8\arcsec, respectively. The first observation revealed a companion at a separation of $1022 \pm 9$ mas, with a position angle (PA) of $238.3 \pm 0.5$ deg, and a flux ratio of $0.0066 \pm 0.0009$ ($\Delta mag = 5.4 \pm 0.2$). In the second observation the companion was detected at a separation of $1006 \pm 9$ mas, PA of $238.6 \pm 0.5$ deg, and flux ratio of $0.0086 \pm 0.0016$ ($\Delta mag = 5.2 \pm 0.3$). The sensitivity curves for both observations are shown in Figure \ref{fig:dirim}. 
\\
With the discovery of the companion to TOI~5786 via speckle imaging, near-infrared adaptive optics imaging was obtained on July 29, 2024, UT with the PHARO instrument \citep{hayward2001} on the Palomar Hale (5m) in the Kcont $(\lambda_o = 2.29; \Delta\lambda = 0.035~\mu$m) and Hcont $(\lambda_o = 1.668; \Delta\lambda = 0.018~\mu$m) filters and on November 11, 2024, UT with NIRC2 on Keck II (10m) in the Jcont $(\lambda_o = 1.2132; \Delta\lambda = 0.0198~\mu$m) filter; both sets of observations were made behind the natural guide star systems \citep{wizinowich2000,dekany2013}. The PHARO pixel scale is $0.025\arcsec$ per pixel and the NIRC2 pixel scale is $0.009942\arcsec$ per pixel. Using a 5-point quincunx dither pattern on Palomar and a 3-point dither pattern on Keck, the reduced science frames were combined into a single mosaiced image with final resolutions of 0.09\arcsec, 0.07\arcsec, and 0.04\arcsec, at Kcont, Hcont, and Jcont respectively. 
\\
The sensitivity of the final combined AO image was determined by injecting simulated sources azimuthally around the primary target every $20^\circ $ at separations of integer multiples of the central source's FWHM \citep{furlan2017}. The brightness of each injected source was scaled until standard aperture photometry detected it with $5\sigma $ significance.  The final $5\sigma $ limit at each separation was determined from the average of all of the determined limits at that separation and the uncertainty on the limit was set by the rms dispersion of the azimuthal slices at a given radial distance.  The Palomar and Keck imaging and sensitivities are shown in (Figure~\ref{fig:dirim}). 
\\
Given the flux ratio in the $I_\mathrm{c}$ band which is closest to the TESS filter we calculate the influence on the derived planet radius of TOI-5786.01. The potential error is less than 0.5~\% and much lower than other error sources (e.g. uncertainty on the stellar radius, scatter of the lightcurve). We therefore do not consider it in our analysis. 
\subsection{High-resolution spectroscopy}
\subsubsection{TRES}
To assess the stellar properties and rule out eclipsing binary scenarios we obtained reconnaissance spectra with the Tillinghast Reflector Echelle Spectrograph (TRES, \cite{furesz2008design}) on the 1.5 m Tillinghast Reflector telescope at the Fred Lawrence Whipple Observatory (FLWO) on Mount Hopkins in Arizona. TRES is a fiber-fed, optical spectrograph with a resolving power of R = 44,000 covering the wavelengths between 390 and 910 nm. We obtained four spectra for TOI-5108 and two for TOI-5786. For one of the TOI-5108 spectra, the observations were interrupted by clouds. All spectra were visually inspected to check for composite spectra and then a multi-order spectral analysis was performed following procedures outlined in \cite{buchhave2010} and \cite{Quinn2012}. Essentially, the spectra were cross-correlated order-by-order against a median combined template to derive radial velocities. Then the spectra are corrected for zero point offsets using a history of standard star observations. We found no signs of composite spectra or larger RV variations indicative of an eclipsing binary.
\\
\subsubsection{SOPHIE}
To establish the planetary nature of the transit events, measure the
masses of the companions, and constrain their orbits, in particular eccentricities,
we observed both stars with SOPHIE. This stabilized \'echelle spectrograph is
dedicated to high-precision radial-velocity measurements in optical wavelengths
on the 193-cm Telescope at the Observatoire de Haute-Provence, France \citep{Perruchot2008,Bouchy2009}. It was already used to characterize several planetary candidates revealed by TESS \citep[e.g. ][]{Moutou2021,konig2022,Martioli2023,bell2024}. We used the high-resolution mode
of SOPHIE (resolving power $R=75,000$) and the fast reading mode of its CCD.
\\
For TOI-5108 we secured 30 observations between February 2022 and March 2024.
We excluded two of them due to low signal-to-noise ratio. For the 28 remaining exposures,
the signal-to-noise ratio per pixel at 550~nm ranges from 28 to 52 (typically 40),
for exposure times between 7 and 35~minutes (typically 15 minutes) depending on the
weather conditions.
\\
For TOI-5786, 15 observations were secured between April 2023 and March 2024.
Signal-to-noise ratios per pixel and exposure times respectively range from
36 to 50 (typically 40) and from 12 to 36 minutes (typically 18 minutes), whereas
one exposure was rejected due to its low signal-to-noise.
\\
The radial velocities were extracted with the standard SOPHIE pipeline using cross
correlation functions (CCF) as presented by \cite{Bouchy2009} and refined by \cite{Heidari2024}. This includes in particular corrections for CCD charge transfer inefficiency, instrumental drifts, and pollution by Moonlight. The derived radial velocities have uncertainties of the order of 3 and 6 m/s, respectively for TOI-5108 and TOI-5786. For both objects, the radial velocities show significant variations in phase with the \textit{TESS} ephemeris. The amplitudes of those variations indicate companions with masses in the planetary range. Additionally, the bisectors of the CCF were computed, and they show no correlations with the radial-velocity variations, as it might be the case if the transits are not due to planets but blended eclipsing binaries instead. The bisectors also constitute an activity indicator (see Sect. \ref{stellarrot}).

\subsubsection{MaHPS}
Together with the SOPHIE observations we also collected spectra with the Manfred Hirt Planet finder Spectrograph \citep[MaHPS, ][]{Kellermann2015,Kellermann2016} consisting of the fiber-fed, high-resolution ($R \sim 60,000$), optical ($385 - 885$ nm) echelle spectrograph FOCES \citep{Pfeiffer1998} and a Menlo Systems Astro Frequency Comb as a wavelength calibration source. 
\\
MaHPS is mounted on the 2.1\,m telescope at the Wendelstein Observatory in southern Germany. It is temperature- and pressure-stabilized down to $<0.01$ K and $<0.1$ hPa inside an aluminum tank that is kept at an artificial overpressure \citep{Fahrenschon2020}. For the wavelength calibration, a Laser Frequency Comb (LFC) from Menlo Systems is used which produces narrow, regularly spaced emission lines between $470$ and $740$ nm. Two fibers are available with one recording the science light, while the other one is used to record the LFC spectrum for simultaneous wavelength calibration to correct for instrumental drifts during exposures \citep{Brucalassi2016,Wang2017,Kellermann2019}. Additionally, spectra of a Thorium-Argon (ThAr) lamp are recorded once every day before science observations for initial wavelength calibration. This calibration can also be used in case the precision of the comb is not needed or not available.
\\
We acquired 141 MaHPS spectra on 48 observing nights between November 2022 and March 2024 for TOI-5108 and 113 spectra on 48 nights between May 2023 and April 2024 for TOI-5786.
If possible we collected three 1800s exposures per night however, for some nights worsening weather conditions or higher priority observations meant we could only take one or two exposures. The observations in this paper belong to the first batch of science observations from the MaHPS spectrograph \citep[see also][]{Ehrhardt2024}. Therefore the limiting observing conditions (e.g. weather, moon, seeing) were not well explored yet and some experimentation was done while observing the objects. This caused the number of spectra that were not suitable for precise RV work to be slightly higher compared to established spectrographs. In total, we have 120 spectra on 44 nights for TOI-5108 that are appropriate for precise RV analysis. For TOI-5786 we have 74 spectra on 35 nights. The first criterion for the evaluation of the spectra was the signal-to-noise ratio (SNR). We excluded data with an SNR less than 50~\% of the mean SNR for the target which was mostly data affected by clouds. There were also cases were the filter of the LFC light was set to the wrong value (often after maintenance work on the LFC) which caused the LFC light to bleed into the target spectrum and introduced shifts in the RV. Lastly, there were observations where the targets were too close to the moon which caused contamination of the spectrum. In all of these cases, we excluded the affected observations from the analysis.  
\\
Additionally, there are nights for both stars where the frequency comb was down for maintenance or the comb flux was too low to be used for wavelength calibration. In both cases, only the wavelength for calibration from the ThAr lamp is used.
\\
We have developed two in-house software packages (GAMSE and MARMOT) to reduce the MaHPS spectra and calculate the radial velocities of the star. GAMSE \citep{Wang2016} is used to reduce the raw data and extract 1D spectra. It also performs a preliminary wavelength calibration based on the ThAr spectra from the same day. MARMOT \citep{Kellermann2020} then uses the 1D spectra and re-calibrates the wavelength solution based on the simultaneous LFC spectra. Next, it performs the barycenter correction using the Python package barycorrpy \citep{kanodia2018python} and masks any cosmic rays left in the spectrum. For the computation of the final RV shifts, MARMOT offers computation via the CCF method as well as the template matching method using a $\chi^2$ fit. 
\\
Stellar templates for the RV calculation in MARMOT are created from the individual observations through the simultaneous fitting of a cubic B-spline function. A separate template is created for each echelle order. First, the input spectra are scaled with a 2nd order polynomial and shifted by applying a multiplicative scaling to their wavelength axis to correct for shifts between the individual spectra and match them. In the following step, the $\chi^2$ statistic of the data points and the B-spline is minimized using the Python program {\tt iminuit} \citep{iminuit} which is a python wrapper of the minuit minimizer. The final result is stored as a function to be evaluated during RV evaluation using either a CCF or $\chi^2$ fit.
\\
For the templates of both stars, we used all observations that were left after the initial bad spectra rejection. We then derived the radial velocities for each order of the individual observations using the $\chi^2$ fitting routine in MARMOT.
\\
Before calculating the final radial velocities for each observation, we apply two corrections for known systematics in the MaHPS spectra. 
\\
The first known systematic offset in the MaHPS spectra are variations of the nightly zero-point (NZP) of the instrument. These variations have been common for many spectrographs and are often caused by instrumental events like detector or fiber changes \citep[see ][]{trifonov2020public}, fiber coupling problems, or thermal effects. For MaHPS the fiber coupling to the telescope is currently still in a provisional state which can cause instabilities of the NZP. The NZP variations also seem to be correlated with sudden changes in the outside temperature which could be propagated through the coupling of the science fiber to the telescope. 
\\
To correct for these nightly zero-point variations we observe one RV standard star between each TOI observation run. In total, we use four reference stars that are observable at different times throughout the year: HD185144, HD9407, HD221354, and HD89269. To build a nightly zero-point master that can be used to correct the science observations we follow an approach that has been used for multiple other spectrographs \citep[e.g. ][]{courcol2015sophie,trifonov2018carmenes,tal2019correcting}. If we had multiple observations of a standard star for one night we first binned them to avoid overweighting a particular star. Then we normalize the radial velocity measurements of each standard star by subtracting the error-weighted average of all their RV values. Lastly, the instrumental NZP is calculated by taking the error-weighted average RV of all RV standard stars that were observed on a given night.
\\
In the case that we do not have the LFC available to correct for instrumental drifts during the night we slightly alter our NZP calculation. Instead of calculating one zero-point for the whole night under the assumption that it should not change over the course of the observations, we use the reference star observations before and after each of the science exposures to correct for both the nightly zero point and the instrumental drifts during the exposures. In these cases, we interpolate between the reference star exposure before and after the science exposures and calculate a zero-point value for each exposure individually. We have found that this improves the precision of our RV measurements substantially for nights without the LFC especially if there are strong variations in the temperature during the night.
For the correction of the RVs of TOI-5108 and TOI-5786, the NZP value is subtracted from the RV value of each order for the respective nights. 
\\
A second systematic can be introduced during the template creation. Since the template is fitted for each order separately and the individual spectra are allowed to shift in wavelength, the templates of individual orders can have small offsets between each other. This can have an impact during the calculation of the final RV for the observations as we weigh the RV of each order by its error which can unevenly amplify the effect of the template offsets. Therefore, after subtracting the NZP from the RV value of each order we normalize the orders to get rid of the offset by subtracting their error weighted average. 
\\
After applying both corrections to the order-wise RVs we calculated the final radial velocities for each observation as the error-weighted average of the RVs from all spectral orders of the observation. The error of the radial velocity values is given by the weighted rms \citep[see ][]{zechmeister2018spectrum}.
\\
Both the MaHPS and SOPHIE RVs that are used in the following analysis are available at the CDS, containing the following information. Column 1 lists the time of the observations, Column 2 gives the derived RV values, Column 3 gives the error of the RV and Column 4 lists the name of the instrument.
\\
In addition to the RVs we compute activity indices from the calcium infrared triplet (IRT) lines following the definition of \cite{kurster2003}. We only use the first two lines of the IRT as the last line is located in the order that is cut off by the upper edge of the detector. 

\section{Stellar analysis} \label{stellar}
\subsection{Spectroscopic stellar parameters} \label{stellarspec}
To characterize the two stellar hosts, we first extract spectroscopic stellar parameters from the observed TRES, MaHPS, and SOPHIE spectra. 
\\
Stellar parameters from the TRES spectra are obtained by cross-correlating the spectra against Kurucz atmospheric models \citep{kurucz1992} to derive stellar effective temperature, surface gravity, metallicity, and rotational velocity using the Stellar Parameter Classification tool \citep[SPC; ][]{buchhave2012,buchhave2014}. We exclude the spectrum of TOI-5108 that was interrupted by clouds for the calculation of the average parameters. The average stellar parameters from the TRES spectra are shown in Table \ref{tab:specana}. SPC also derives the rotational velocities for our two stars: $v \sin i =4.1 \pm 0.5$ km/s for TOI-5108 and $v \sin i=9.0 \pm 0.5$ km/s for TOI-5786.
\\
For the MaHPS spectra, we compare the stellar templates that we created from our observations to a library of stars with precise and accurate T$_{\text{eff}}$, $\log g$ and $\text{[Fe/H]}$ values using the Python code {\tt SpecMatch-Emp} \citep{yee2017}. First, we normalize our MaHPS template and then fit it order-wise to the 404 library spectra in the wavelength range between 500 nm and 640 nm. The final parameters for our two stars are determined from a linear combination of the five best-fitting library spectra. In the case of TOI-5786, one of the five best-fitting stars had a significantly higher radius (2.2 R$_\odot$ vs 1.2 - 1.5 R$_\odot$) with a lower temperature (5600~K vs 6000 - 6300~K). For the linear combination of reference spectra, we remove this star and use the sixth best-fitting reference instead. While {\tt SpecMatch-Emp} does fit for the projected rotational velocity $v \sin i$ and does not include it in the final parameters, the reference stars often have measured $v \sin i$ values. For TOI-5108 the $v \sin i$ of the reference stars is between 2.3 and 4.2 km/s and thus close to the value derived from the TRES spectra. For TOI-5786 four of the reference stars have $v \sin i$ between 6.6 and 7.4 km/s with one outlier at a $v \sin i$ of only 1.2 km/s.
\\
For the SOPHIE spectra, we selected the ones without moonlight pollution and constructed an average spectrum for both stars. Following the method presented by \cite{Santos2013} and \cite{Sousa2021}, we measured on each averaged spectrum the equivalent widths of selected iron lines, and derived spectral parameters from them using the ARES+MOOG, including $\text{T}_{\text{eff}}$, $\log g$, and $\text{[Fe/H]}$. The vsini values derived from SOPHIE are $3.3 \pm 1.0$ km/s and $7.0 \pm 1.0$ km/s, for TOI-5108 and TOI-5786, respectively. They were computed from the CCF parameters, using the calibrations of \cite{Boisse2010}.
\\
The derived spectroscopic parameters are listed in Table \ref{tab:specana}. We use the weighted average of the values from the three spectrographs as priors on the spectroscopic stellar parameters in the following analysis to derive the bulk parameters of our two target stars.  

\begin{table}
    \centering
    \begin{threeparttable}[b]
    \centering
    \caption{\label{tab:specana}Stellar parameters from the analysis of the TRES, MaHPS, and SOPHIE spectra.}
    \begin{tabular}{lccc}
    \hline
    \hline
      Parameters   & TRES & MaHPS & SOPHIE \\
      \hline
      \\
      TOI-5108  \\
      \\
      $\text{T}_{\text{eff}}$ [K]& $5816 \pm 50$  & $5763 \pm 110$ & $5880 \pm 62$ \\
      $\log g$& $4.2 \pm 0.10$ & $4.16 \pm 0.12$& $4.20 \pm 0.11$\\
      $\text{[Fe/H]}$\tnote{a)} & $0.19 \pm 0.08$ & $0.19 \pm 0.09$ & $0.19 \pm 0.05$ \\
      \\
      TOI-5786 \\
      \\
       $\text{T}_{\text{eff}}$ [K]& $6162 \pm 51$ &$6175 \pm 110$ & $6392 \pm 72$ \\
        $\log g$ & $4.20 \pm 0.10$ & $4.26 \pm 0.12$ & $4.2 \pm 0.1$\\
        $\text{[Fe/H]}$\tnote{a)}&$0.09 \pm 0.08$ & $0.09 \pm 0.09$ & $0.18 \pm 0.05$ \\
    \hline
    \end{tabular}
        \begin{tablenotes}
       \item [a)] The TRES SPC analysis technically measures [M/H] but it is similar enough to [Fe/H] to allow for this comparison.
     \end{tablenotes}
     \end{threeparttable}
\end{table}
\subsection{Bulk stellar parameters} \label{stellarsed}
We then derive the bulk stellar parameters of both stars by analyzing their broadband spectral energy distribution (SED) with the Python package ARIADNE \citep{ariadne2022}. ARIADNE uses Bayesian model averaging to combine the results of fitting different stellar atmosphere models to account for and reduce model-specific systemic biases in the final parameters. For both stars, we use photometric data from the filters Johnson B, V, Gaia DR3 G, RP and BP, 2MASS JHK, and ALL-WISE W1 and W2 (W3 and W4 are not used as the PHOENIX V2 grid does not extend to that wavelength range) and include the GALEX NUV and FUV data for TOI-5108. 
\\
We convolve stellar atmosphere models from PHOENIX V2 \citep{husser2013}, BT-Settl, BT-NextGen, BT-Cond \citep{Hauschildt1999,allard2012}, Castelli \& Kurucz \citep{Castelli2003}, and Kurucz \citep{Kurucz1993} with the chosen filter bandpasses and interpolate them in T$_{\text{eff}}$ - $\log g$ - [Fe/H] to fit to the photometric data. 
\\
The model for the SED includes six main parameters: the effective temperature T$_{\text{eff}}$, the surface gravity of the star log g, the metallicity [Fe/H], the extinction $\text{A}_v$ and the stellar radius $R_*$ and distance $D$ via the dilution factor $(R_*/D)^2$ which is used to obtain the observed fluxes from the theoretical grid fluxes \citep{cassegrande2014}. As an additional free parameter, an excess noise term $\epsilon_i$ for each photometric measurement is included.
\\
We use a normally distributed prior around the Bailer-Jones distance estimate from Gaia EDR3 \citep{Bailer-Jones2021} for the distance. The stellar radius and extinction are given flat priors ranging from 0.05 to 20 R$_\odot$ for the radius and 0 to the maximum line-of-sight extinction calculated from the updated SFD galactic
dust map \citep{Schlegel1998,Schlafly2011} through the dustmaps python package \citep{Green2018}. For the parameters T$_{\text{eff}}$, $\log g$ and $\text{[Fe/H]}$ we use the results from our spectroscopic analysis in Section \ref{stellarspec} to impose Gaussian priors centered around the weighted average of the derived values.
\\
Each stellar atmosphere model is fitted to the data separately and then the final values for all parameters are computed using a weighted average where the weights are given by the relative probability of the respective models. In the last step, the mass and age of the stars are estimated by comparing interpolated MESA Isochrones and Stellar Tracks \citep[MIST,][]{Dotter2016,Choi2016} evolutionary models to the parameters derived from the SED fit using the {\tt isochrones} \citep{Morton2015} software package. 
\\
As an independent confirmation of the stellar parameters we use the {\tt isoclassify} package \citep{Huber2017} to fit interpolated MIST and PARSEC \citep{Bressan2012} isochrones directly to the broadband photometry and our derived spectroscopic parameters to model R$_*$, M$_*$, $\text{A}_v$, L$_*$ and the stellar age. {\tt isoclassify} has recently been used to homogenously derive a large sample of stellar properties from Kepler, K2, and \textit{TESS} stars \citep{Berger2020,Berger2023}. We used the GAIA and 2MASS bands and imposed the same priors on the spectroscopic parameters that were used in the SED fitting. We first derive the luminosity directly from the bolometric flux and the GAIA parallax using the Bayestar19 \citep{Green2019} reddening map for the extinction correction. This luminosity together with the spectroscopic constraints on T$_{\text{eff}}$, $\log g$ and $\text{[Fe/H]}$ is used to fit a grid of interpolated isochrones to derive the bulk stellar parameters. We perform this analysis for both the MIST and PARSEC isochrones.
\\
The results of the SED (see Figure \ref{fig:sed}) and the {\tt isoclassify} fitting are shown in Table \ref{tab:starcomp}. All three values agree within the errorbars. For the final stellar parameters in Table \ref{tab:star} we adopt the values from the {\tt isoclassify} analysis with MIST isochrones as this approach was also used in a homogenous analysis of the TESS-Keck survey \citep{MacDougall2023}. 
\\
For the projected rotational velocity we adopt the values from SOPHIE spectra which are also consistent with the expected values from the {\tt SpecMatch-Emp} analysis of the MaHPS spectra. The rotational velocities derived from TRES are slightly higher, however, since the SPC does not take into account macroturbulence they can be slightly overestimated.
\begin{table*}
    \centering
    \begin{threeparttable}[b]
    \centering
    \caption{\label{tab:star}Stellar parameters of TOI 5108 and TOI 5786}
    \begin{tabular}{lccc}
    \hline
    \hline
      Parameters   & TOI 5108 & TOI 5786 & Reference \\
      \hline
      \\
      Identifiers  \\
      TIC ID & TIC 350348197 & TIC 40292751 & TICv8\\
      2MASS ID & J11050459+1114476 & J19331736+3042152& 2MASS\\
      GAIA ID & 3868491612036282752 & 2032764938419327616 & \textit{Gaia} DR3 \\
      \\
      Astrometric Properties \\
        $\alpha$ (J2000)  & 11:05:04.61 &19:33:17.36 & \textit{Gaia} DR3 \\
        $\delta$ (J2000) & 11:14:47.78 & 30:42:15.28 & \textit{Gaia} DR3 \\
        Parallax (mas) & $5.160 \pm 0.013$ & $7.6354 \pm 0.03$ & \textit{Gaia} DR3\\
        RUWE & 0.917 & 0.841 & \textit{Gaia} DR3 \\
        \\
        Photometric Properties\\
        \textit{TESS} (mag) & $9.185 \pm 0.007$& $9.773 \pm 0.007$ & TICv8\\
        FUV (mag) & $22.3 \pm 0.5$& -- & GALEX\\
        NUV (mag) & $15.200 \pm 0.009$ & -- & GALEX\\
        B (mag) & $10.373 \pm 0.038$& $10.766 \pm 0.068$ & TICv8.2\\
        V (mag) & $9.75 \pm 0.03$ & $10.186  \pm 0.005$ & TICv8.2\\
        Gaia$_{BP}$ (mag) & $9.928 \pm 0.003$ & $10.402 \pm 0.003$ & \textit{Gaia} DR3\\
        Gaia (mag) & $9.612 \pm 0.003$ & $10.142 \pm 0.003$ & \textit{Gaia} DR3\\
        Gaia$_{RP}$ (mag) & $9.132 \pm 0.004$ & $9.715 \pm 0.004$ & \textit{Gaia} DR3\\
        J (mag) & $8.61 \pm 0.02$ & $9.26 \pm 0.03$ & 2MASS\\
        H (mag) & $8.40 \pm 0.04$ & $9.04 \pm 0.03$ & 2MASS\\
        K$_S$ (mag) & $8.30 \pm 0.04$ & $8.96 \pm 0.02$ & 2MASS\\
        W$_1$ (mag) & $8.23 \pm 0.03$ & $8.95 \pm 0.03$ & WISE\\
        W$_2$ (mag) & $8.29 \pm 0.02$ & $8.96 \pm 0.03$ & WISE\\
        W$_3$  (mag) & $8.25 \pm 0.03$ & $8.92 \pm 0.03$ & WISE\\
        W$_4$  (mag) & $8.3 \pm 0.4$ & $9.0 \pm 0.5$ & WISE\\
        \\
        Derived Properties
        \\
        T$_{\text{eff}}$ [K]& $5808 \pm 40$ & $6235 \pm 50$ & This work\\
        $\log g$ [cgs]& $4.26^{+0.05}_{-0.04}$  & $4.26^{+0.04}_{-0.03}$ & This work\\ 
        $\left[\textrm{Fe/H}\right]$ [dex] & $0.20^{+0.02}_{-0.05}$  & $0.08^{+0.1}_{-0.08}$ & This work\\
        R$_*$ [R$_\odot$] &  $1.29^{+0.04}_{-0.04}$ & $1.36^{+0.03}_{-0.05}$ & This work \\ 
        M$_*$ [M$_\odot$] & $1.10^{+0.03}_{-0.03}$  & $1.23^{+0.04}_{-0.03}$ & This work\\ 
        $v \sin i$ [km/s] & $3.3 \pm 1.0$& $7.0 \pm 1.0 $& This work\\
       Age [Gyr] & $5.8^{+1.0}_{-0.8}$ & $2.5^{+0.5}_{-0.7}$ & This work\\ 
       Distance [pc] & $131.1 \pm 0.6$ & $193.8 \pm 0.5$ & This work \\
       \hline
    \end{tabular}
    \begin{tablenotes}
       \item References: \textit{Gaia} DR3 \citep{Gaia2020}; 2MASS \citep{2mass2003}; TICv8 \citep{tic2022}; GALEX \citep{galex2005}; WISE \citep{wise2010}.
       \end{tablenotes}
     \end{threeparttable}
\end{table*}
\subsection{Stellar rotation period} \label{stellarrot}
\begin{figure}
    \centering
    \includegraphics[width=0.9\columnwidth]{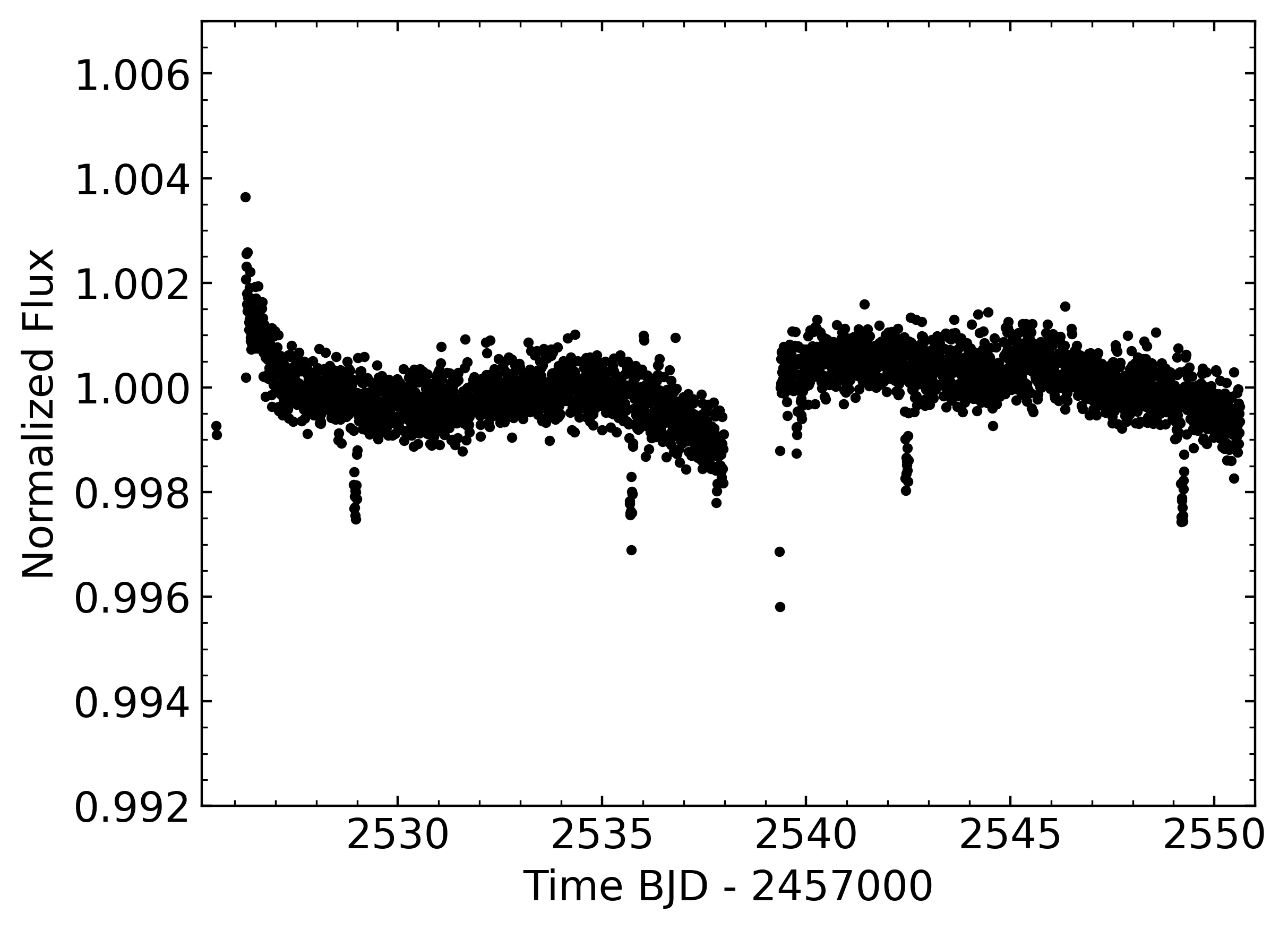} 
    \includegraphics[width=0.9\columnwidth]{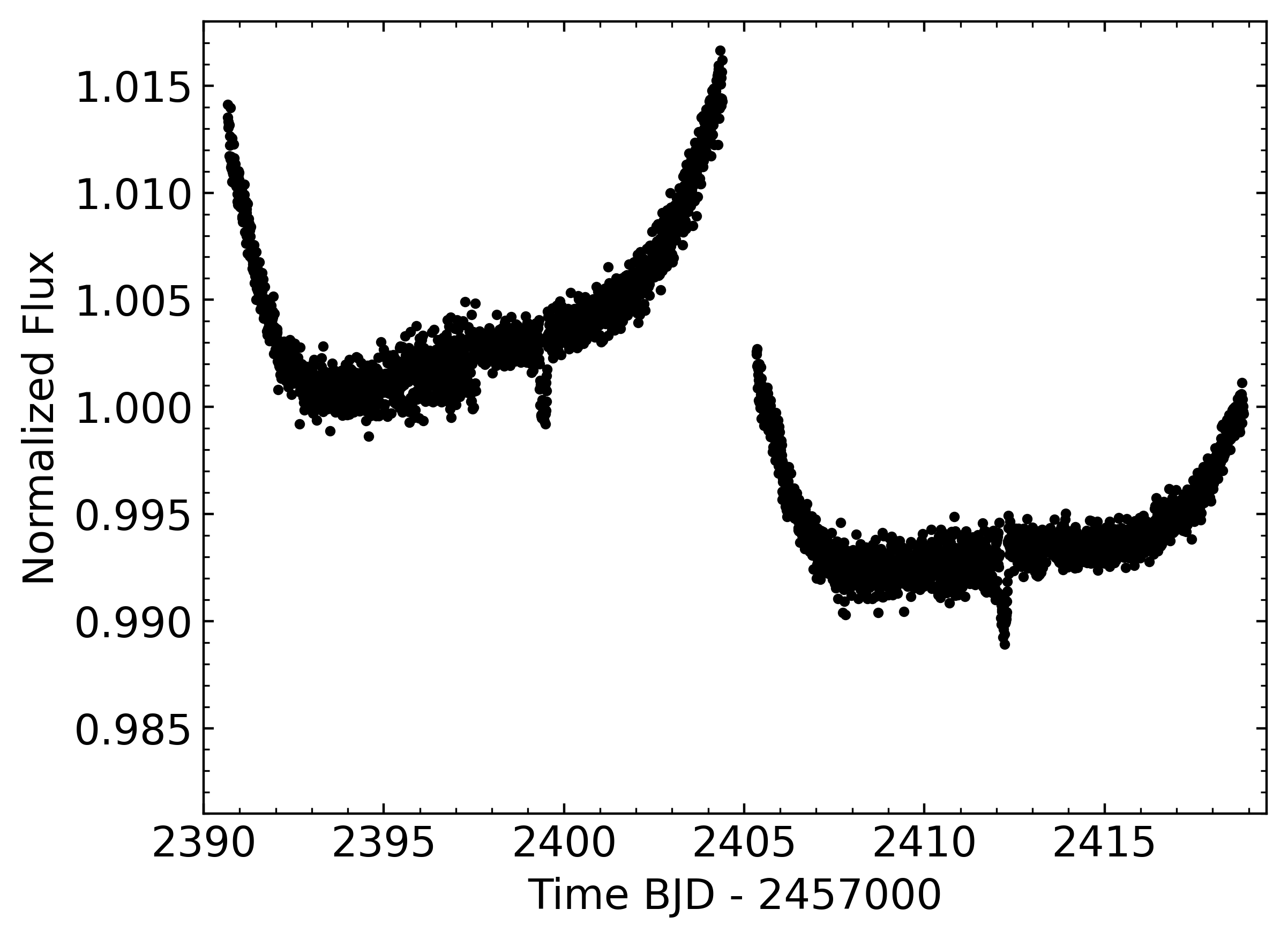} 
    \caption{SAP flux of TOI-5108 and TOI-5786. The lower plot shows the lightcurve of TOI-5786 which has strong systematics at the edges and an offset between the two sectors. For TOI-5108 (upper plot) the systematics are less severe. }
    \label{fig:sapfluxes}
\end{figure}

\begin{figure*}
    \centering
    \includegraphics[width=1.9\columnwidth]{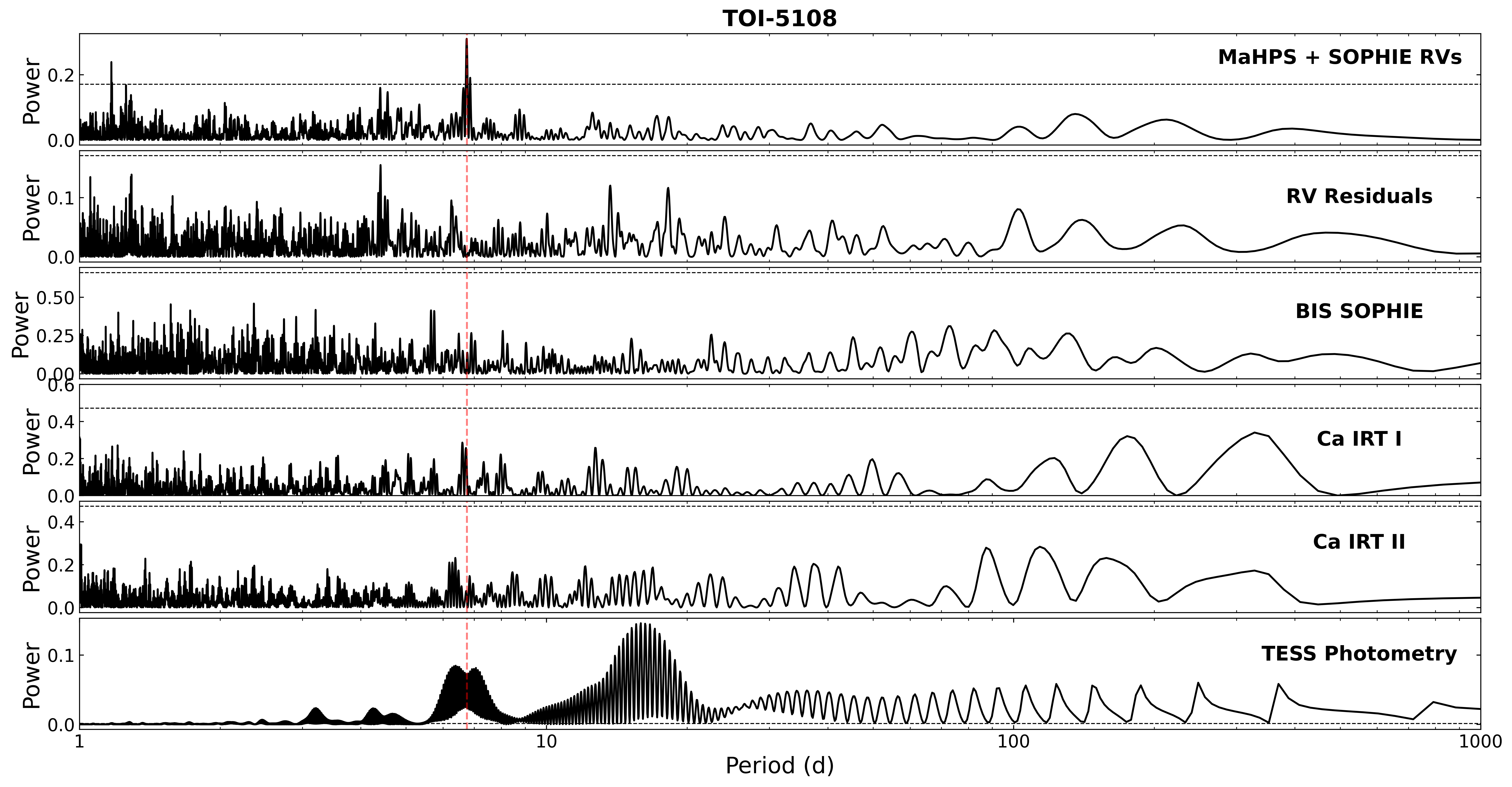} 
    \includegraphics[width=1.9\columnwidth]{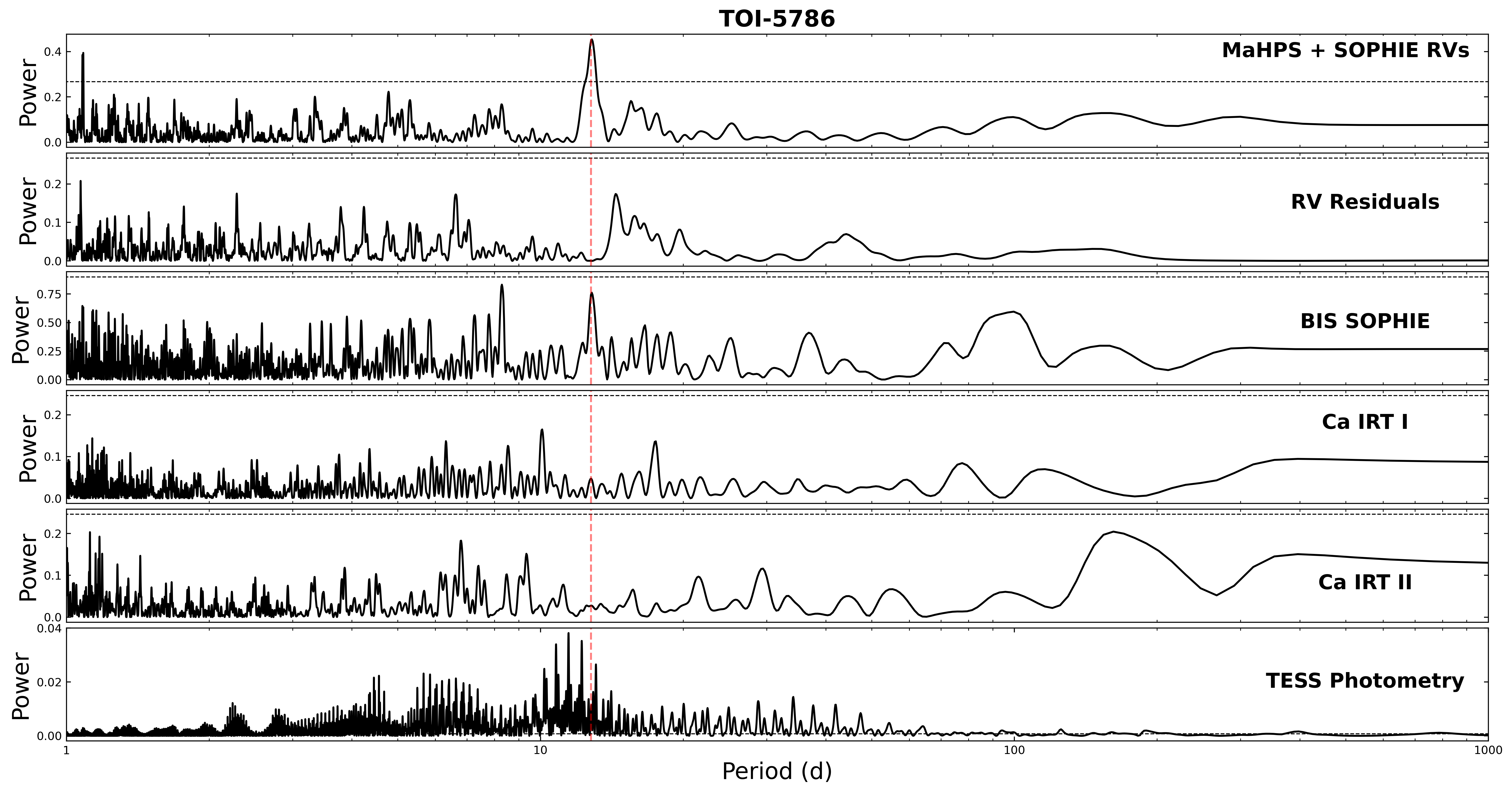} 
    \caption{GLS periodograms of TOI-5108 and TOI-5786. The top periodograms belong to TOI-5108. The first and second rows are the RV data from MaHPS and SOPHIE before and after subtracting the signal of the transiting planet. The next three rows show the periodograms of the activity indicators derived from the SOPHIE spectra (BIS) and MaHPS spectra (Ca IRT). The last row shows the periodogram of the \textit{TESS} data. The dotted red line shows the period of the transiting planet and the dotted black line is located at the $0.1\,\%$ FAP. Bottom periodograms are the same as the top ones but for TOI-5786.}
    \label{fig:glspg}
\end{figure*}

From the projected rotational velocity $v \sin i$, we can calculate an estimate for the rotational period of the target stars. Given the stellar radii from Table \ref{tab:star} we get $\text{P}_{rot}/\sin i  = 20.0 \pm 6.0$ days for TOI-5108 and $\text{P}_{rot}/\sin i = 9.8 \pm 1.4$ days for TOI-5786. In theory, these are strict upper limits since the rotational velocity could be larger given an inclination of the stellar rotation axis $<90^\circ$, however, measuring the rotational velocity of a star from the broadened spectra can be difficult as many other effects can broaden spectral lines (e.g. activity, turbulence) which are often difficult to model for the specific star. Underestimating these effects in the modeling can lead to an overestimation of the $v \sin i$ measured from the spectra translating into a larger range of possible rotational periods.
\\
To further investigate the rotational period of the two stars, we construct Generalized-Lomb-Scargle \citep[GLS, ][]{Zechmeister2009} periodograms from our RV data to search for any periodicities in the data that are not associated with our planetary signals. For both stars, the peaks with the highest power in the periodogram are associated with the period of the transiting planet. Therefore, we use the best-fit values of the periodogram to subtract the planetary signal. Figure \ref{fig:glspg} shows the periodogram of the RVs and the residuals. After subtracting the planetary signal there are no more signals above the false-alarm probability (FAP) threshold of $0.1\%$.
\\
Additionally, we investigate the bisector of the CCF of the SOPHIE spectra and the first two lines of the calcium infrared triplet derived from the MaHPS spectra. We find no strong signals above a $0.1\%$ FAP. Even below the FAP threshold, there are no distinct signals that are common in the RV residuals and the activity indicators. For TOI-5108 we find a peak close to the chosen FAP level in the RV residuals at 4.4 days and a couple of peaks in the two calcium lines at longer periods above 100 days. For TOI-5786 there are two distinct peaks in the SOPHIE bisector, one at the period of the planet and one at 8.2 days which is close to the estimated rotational period from $v \sin i$. 
\\
Lastly, we check the \textit{TESS} lightcurves for signals of stellar activity. Due to the 2:1 resonance orbit of \textit{TESS} and the Earth-Moon system, observations are subject to significant contamination from Earthshine and moonlight over the 13.7-day period. This effect and other systematics can severely hinder the detection of astrophysical signals. However, the applied corrections from SPOC can sometimes remove stellar signals in addition to instrumental systematics as both can occur on similar timescales. For cases where the expected rotational period is larger than 13.7 days, it can be better to use the SAP lightcurves to ensure that no stellar signals have been removed. However, this is only sensible if the lightcurves are not dominated by instrumental effects masking all stellar signals. 
\\
As the expected rotation period of TOI-5108 is larger than 13.7 days and the SAP lightcurve is not dominated by systematics (see Figure \ref{fig:sapfluxes}) we use it for the periodogram analysis.
For TOI-5786, the SAP photometry shows large instrumental effects and offsets for all sectors (see Figure \ref{fig:sapfluxes}). Given the large systematics in the SAP lightcurve and the fact that we expect the rotation period of TOI-5786 to be shorter than 13.7 days, we use the PDSCAP flux for the analysis. Since sectors 14 and 54 have large gaps in their PDCSAP data (see Figure \ref{fig:tesslc}) we only use sectors 40, 42, 74, and 81 for the Fourier analysis. In both cases, we mask out the transits before the analysis.
The periodograms of the \textit{TESS} data shown in Figure \ref{fig:glspg} are subject to strong aliasing due to the large gaps between the sectors. For TOI-5108 we find a forest of peaks around 15.2 days while for TOI-5786 a similarly strong feature is seen around a period of 11.7 days. As these are close to the 13.7 day period of the spacecraft and there are clearly systematic effects still present in the data which makes the interpretation of these signals difficult. 
\\
We find no common features between the different datasets and also no significant signals in individual periodograms. Therefore, we cannot constrain the stellar rotation period for either star with the available data.
\subsection{Characterization of the TOI-5786 companion star}
We perform a first characterization of the detected companion of TOI-5786 based on the $Jcont$, $Hcont$, and $Kcont$ magnitudes. The magnitude differences of the two stars were measured with simple aperture photometry where the aperture radius was scaled to the FWHM of each image and sky radii were set to values outside the pair of stars.  The magnitude differences are $\Delta Jcont = 4.022\pm 0.013$~mag, $\Delta Hcont = 3.743\pm 0.017$~mag and $\Delta Kcont = 3.413\pm 0.014$~mag; the separation and position angle (EofN) of the companion relative to the primary star are $1.03\pm0.01\arcsec$ and $238.9\pm0.2^\circ$.  
\\	
Using the measured infrared magnitude differences, the unresolved 2MASS magnitudes have been deblended to retrieve real apparent magnitudes and infrared colors for the two stars. Based upon a 2MASS color-color diagram (Figure~\ref{fig:aoimaging}), the infrared colors of the stars are consistent with a late-type F-star ($\sim$ F7V) for TOI-5786 (in agreement with the more detailed spectroscopic analysis of the star) and a mid-type M-star ($\sim$M5V), suggesting that the companion is a bound M-star. With a separation of $1.03\pm0.01\arcsec$ and a distance of $193.8 \pm 0.5$~pc we calculate a projected separation of $199.6 \pm 2.1$~au which is sufficiently wide not to inhibit the 13-day orbital period of the planet.
\begin{figure}
    \centering
    \includegraphics[width=0.9\columnwidth]{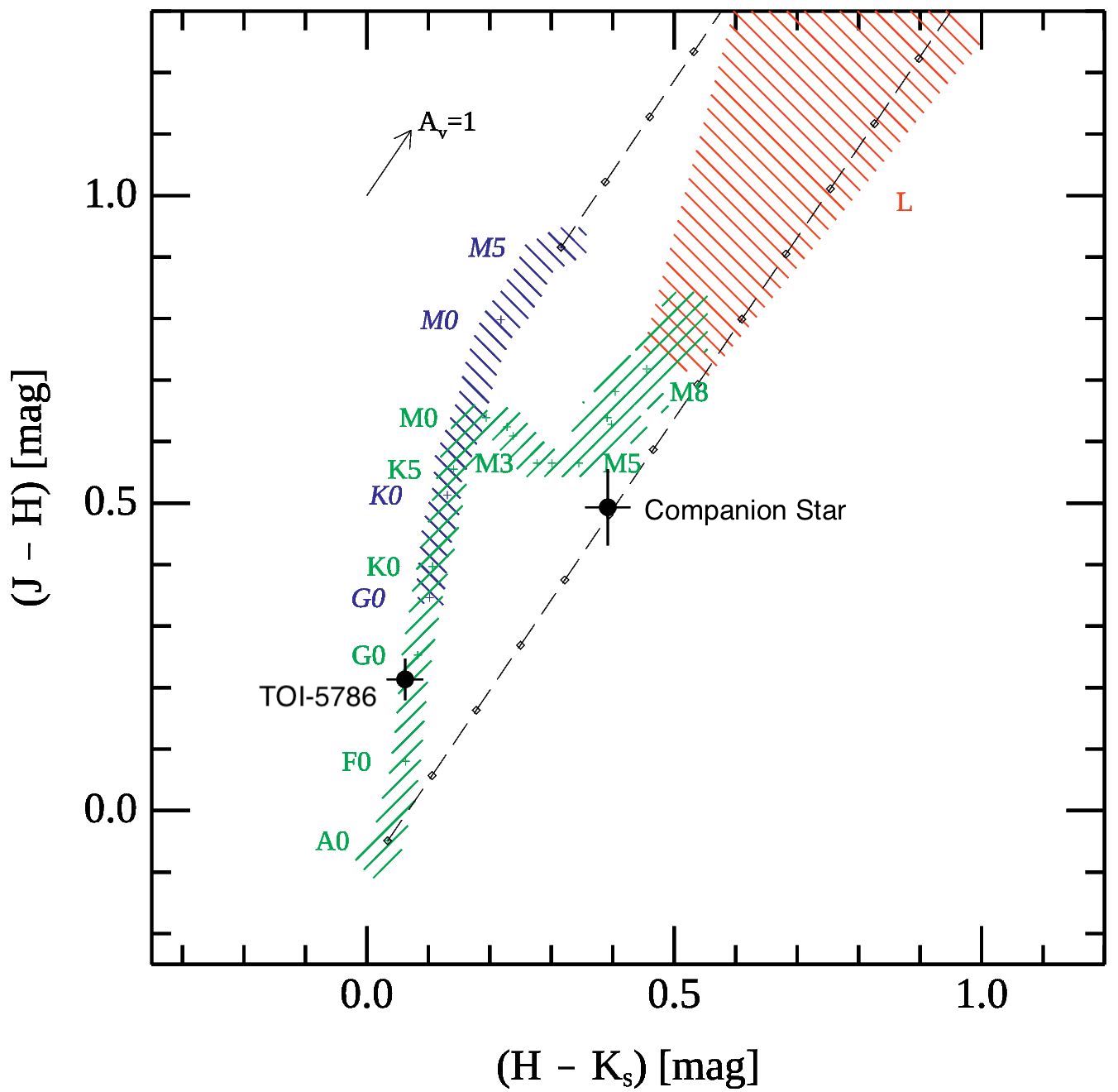}
    \caption{Spectral type classification of TOI-5786 and the detected companion. Using the AO imaging, the 2MASS magnitudes have been deblended and the two stars have been placed on a JHKs color-color diagram.
    }
    \label{fig:palomar_aoimaging}
\end{figure}
\section{Modeling the planetary systems} \label{Analysis}
To derive the orbital and planetary parameters for our planet candidates we perform a joint analysis of our photometric and spectroscopic data using the python package \texttt{juliet} \citep{espinoza2019juliet}. \texttt{juliet} employs the \texttt{batman} package \citep{kreidberg2015batman} to build the transit model for the photometry and \texttt{radvel} \citep{fulton2018radvel} to build the Keplerian model for the RV analysis. For the fitting, \texttt{juliet} uses nested sampling algorithms that perform thorough sampling of the parameter space and at the same time allow for a comparison of different models via the difference in Bayesian evidence $\Delta \ln(z)$. In the following analysis, we used the dynamic nested sampling algorithm implemented via dynesty \citep{speagle2020dynesty} to obtain the posterior distribution of the fitted parameters. 
\\
For both planets, the photometric data is modeled using the mid-transit time $T_0$, the orbital period $P$, the impact parameter $b$, and the planet-to-star radius ratio $R_p/R_*$. Instead of the scaled-semi-major axis $a/R_*$, we include the stellar density $\rho_*$ as a fitting parameter, because it can be constrained from our stellar analysis. For the different photometric instruments we include a jitter term for excess noise $\sigma_i$, a relative flux offset $M_i$ and the quadratic limb-darkening parameters $q_1$ and $q_2$ in the parametrization of \cite{Kipping2013}, which ensures that only values that are physically possible are explored in the sampling. As the reduction for our photometric data already includes a correction of possible contaminating sources we fix the dilution factor of each instrument to 1.
\\
For the radial velocity data, we include a relative offset parameter $\mu_i$ and a jitter term $\omega_i$ for each instrument in addition to the radial velocity amplitude $K$.
\\
For the eccentricity and the argument of periastron we fit for $S_1 = \sqrt{e} \sin \omega$ and $S_2 = \sqrt{e} \cos \omega$ which has shown to allow for fast, yet thorough sampling of the eccentricity \citep{eastman2013exofast}. This parametrization also mitigates the Lucy-Sweeney bias in the eccentricity measurement \citep{Lucy1971}.
\\
We use the stellar parameters derived in Section \ref{stellar} to impose Gaussian priors on both the stellar density and the limb-darkening parameters for the \textit{TESS} lightcurves. The mean of the density prior is calculated from the stellar masses and radii from Table \ref{tab:star} with the derived error as the variance. 
\\
Limb-darkening parameters for our photometric data are calculated using the Synthetic-Photometry/Atmosphere-Model (SPAM) method of \cite{Howarth2011}. In this approach, synthetic lightcurves are created with an accurate representation of the stellar intensity profile through the use of theoretical limb-darkening coefficients derived from the non-linear law and assuming full knowledge of the planetary and stellar properties. These synthetic lightcurves are then fitted with all parameters fixed except for the desired limb-darkening law. This has been shown to minimize the difference between theoretically and empirically derived limb-darkening parameters \citep{Patel2022}.
\\
We use the ExoCTK package \citep{Bourque2021} to compute both, the non-linear limb-darkening coefficients for the synthetic input lightcurves from PHOENIX stellar atmosphere models and the final SPAM coefficients from the quadratic law ($u_1$ and $u_2$) to be used in the fitting of the orbit. The two quadratic limb-darkening coefficients are then transformed to $q_1$ and $q_2$ via the relations given by \cite{Kipping2013}. 
\\
The other parameters are given uniform priors with the period and transit mid-time bounds being relatively tightly centered around the predicted values from the \textit{TESS} lightcurves to be more efficient in the sampling.
\\
For each planet, we perform multiple fits with different models and compare them using the Bayesian evidence ($\ln Z$). A $\Delta \ln Z > 2$ between two models constitutes a moderate preference for one model over the other, while a $\Delta \ln Z > 5$ indicates a strong preference \citep{trotta2008bayes}.
\\
In particular, we are interested in two quantities when comparing models: the eccentricity of the orbit and the presence of other signals either through unaccounted systematics, stellar activity, or other planets in the system.
To investigate whether the data for the two planets warrants an eccentric orbit we perform two separate fits for each planet. In one fit the eccentricity $e$ and the argument of periastron $\omega$ are fixed to $0$ and $90^\circ$ respectively and in the other, both parameters are allowed to vary freely. We then use the Bayesian evidence to compare the two models and determine whether the data warrants an eccentric solution for the orbit. 
\\
Additionally, we attempt to model any remaining systematics and possible effects of stellar activity in our data using Gaussian process regression \citep[GP, see e.g. ][]{Aigrain2023}. For this, \texttt{juliet} employs multiple different GP Kernels via the \texttt{george} \citep{george} and \texttt{celerite} \citep{celerite} packages. While Gaussian process regression can be a powerful tool to recover small planetary signals and measure precise masses in the presence of noise from stellar activity \citep{Barragan2019,Barragan2023,Mantovan2024}, one has to be careful not to overfit the data which can in turn lead to masking or altering of the planetary signals \citep{Ahrer2021,Rajpaul2021,Blunt2023}. Therefore, in the following analysis, we are careful to only apply GP where the data warrant it.  

\subsection{TOI-5108}
\begin{figure*}
    \centering
    \includegraphics[width=1.9\columnwidth]{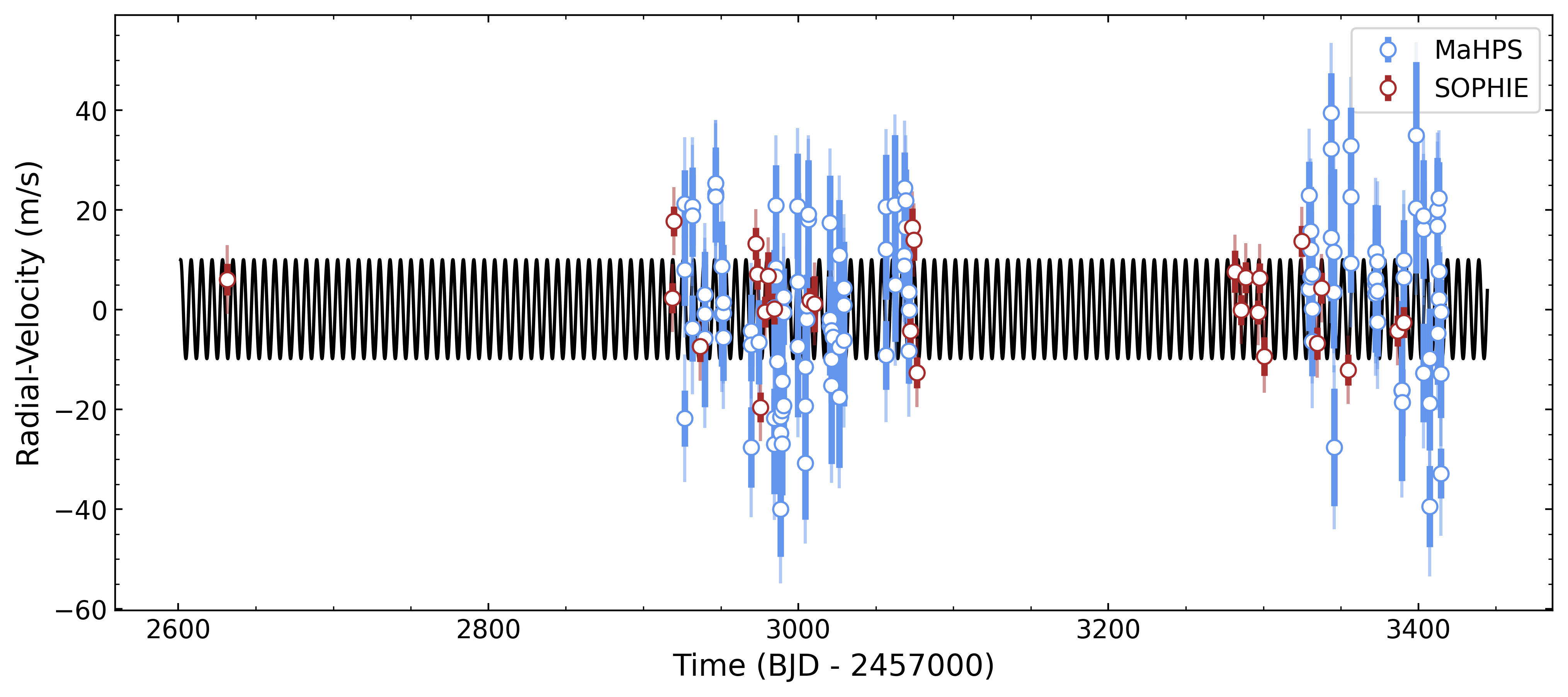} 
    \includegraphics[width=1.9\columnwidth]{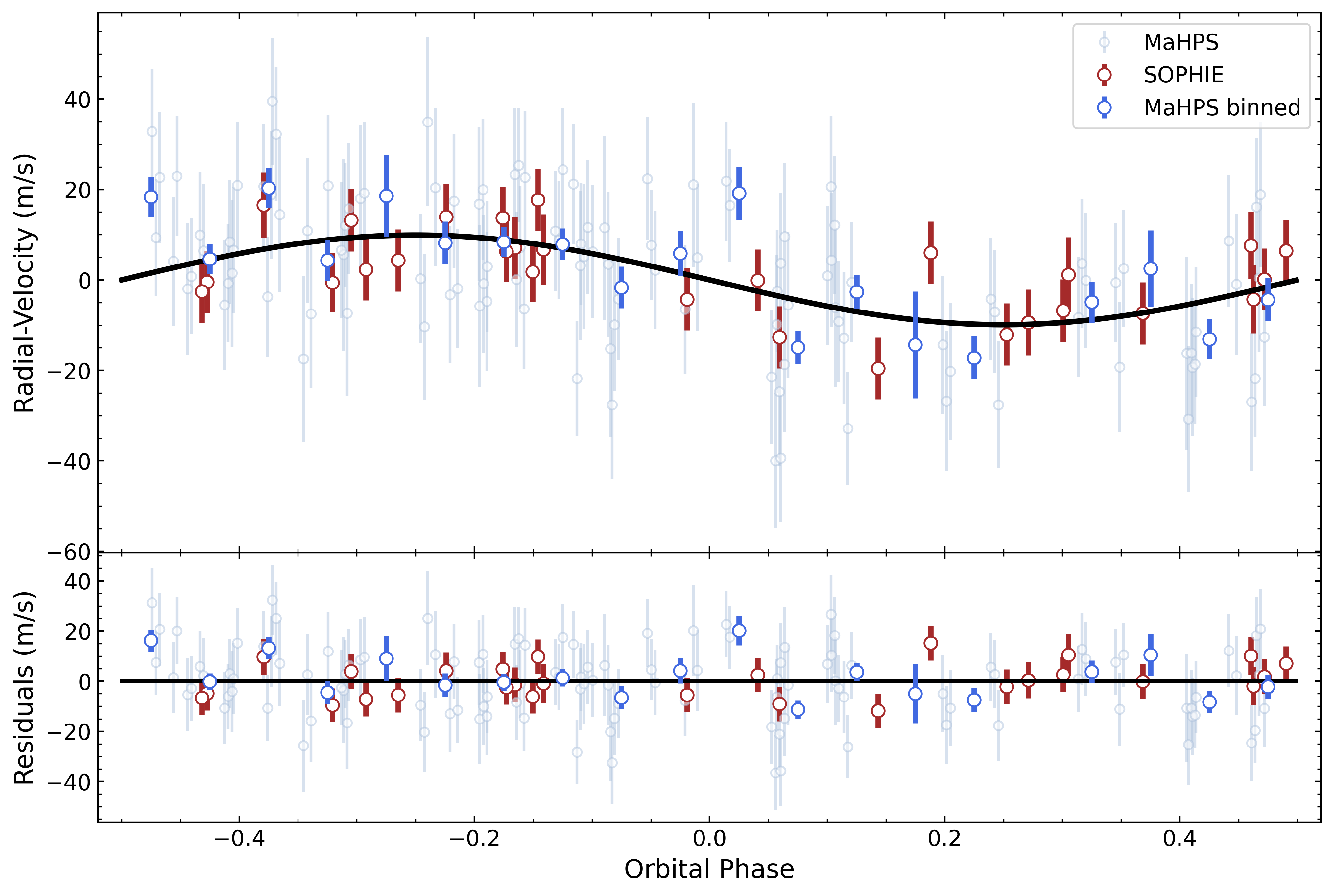} 
    \caption{Upper panel: Radial velocity curve of TOI-5108 with the best fit Keplerian model for TOI-5108\,b in black. Middel panel: Phase-folded radial velocity curve. The dark blue data points represent the MaHPS RV data binned at 0.05 phases. Light blue is the unbinned MaHPS data and red is the SOPHIE data. Lower Panel: Residuals after subtracting the signal of TOI-5108\,b.}
    \label{fig:5108rv}
\end{figure*}

We derive the orbital parameters for the companion of TOI-5108 by using a joint fit of the radial velocities from MaHPS and SOPHIE and the photometry of \textit{TESS} together with the KeplerCam, LCOGT, and MuSCAT2 transits. 
\\
As the SPOC pipeline already detrends the \textit{TESS} data and there are no signs of significant remaining systematics we do not include a GP kernel on the \textit{TESS} data. 
\\
We perform two fits once to a circular and once to an eccentric model for the orbit of TOI-5108\,b on the combined dataset. The difference between the Bayesian log evidence is 7 in favor of the circular model.
\\
Early in the observations, the data showed a slight trend especially in the MaHPS observations but also between the 2022 and 2023 SOPHIE observations (although in 2022 only one observation was taken). The trend got weaker with the addition of the 2024 observations. Nevertheless, we investigate the presence of another planet in our data with another fit where we have broad uniform priors for the period, $T_0$, and the RV amplitude. The two-planet fit is strongly disfavored compared to the one-planet models with a $\Delta \ln Z$ of 7. 
\\
As a consistency check, we also compare the derived RV amplitudes when only one of the spectroscopic datasets is used. The RV amplitude from the MaHPS spectra ($12 \pm 2$~m/s) is consistent with the one derived from SOPHIE ($8.4 \pm 1.9$~m/s).
\\
The final values for the properties of the planet around TOI-5108 adopting the values of the circular model are shown in Table \ref{tab:toi5108} and a plot of the phase-folded RV and time-series are shown in Figure \ref{fig:5108rv}.
\begin{table*}
    \centering
    \begin{threeparttable}[b]
    \centering
    \caption{\label{tab:toi5108}Priors and final parameter values from the \texttt{juliet} fit of TOI-5108.}
    \begin{tabular}{lccl}
    \hline
    \hline
    \\
    Parameter Name & Prior & Derived Values  & Description \\
    \\
    \hline
    \\
    Fitting Parameters \\
    \\
        $P$ [d]        &      $\mathcal{U}[6.7,6.8]$       &  $6.753581 	^{+0.000007}_{-0.000007}$    & Orbital Period\\
        $T_0$ (BJD$_{\rm TDB} -2457000$)   & $\mathcal{U}[2569.1 ,2569.8]$  & $2569.4778  ^{+0.0005}_{-0.0005}$   & Time of Mid-Transit \\
        $R_p / R_*$ & $\mathcal{U}[0.001, 1]$ & $0.0472^{+0.0007}_{-0.0007}$  &  Planet-to-star radius ratio\\
        $b$ & $\mathcal{U}[0, 1]$& $0.871^{+0.009}_{-0.010}$ &  Impact parameter \\
        $\sqrt{e} \sin \omega$ & fixed & 0  & Parametrization of $e$ and $\omega$ \\
        $\sqrt{e} \cos \omega$ & fixed& 0  & Parametrization of $e$ and $\omega$\\
        $K$ [m~s$^{-1}$]      & $\mathcal{U}[0,40]$       &  $9.9^{+1.4}_{-1.3}$  & RV Amplitude\\
        $\rho_*$ [kg/$\text{m}^3$]& $\mathcal{N}[720,90]$ & $745^{+80}_{-70}$ & Stellar Density\\
        \\
        Instrumental Parameters \\
        \\
        $q_{1, TESS}$ & $\mathcal{N}[0.28,0.1]$ & $0.26^{+0.06}_{-0.06}$ & limb-darkening parameter\\
        $q_{2, TESS}$ & $\mathcal{N}[0.33,0.1]$ & $0.40^{+0.10}_{-0.1}$ & limb-darkening parameter\\
        $\mu_{\rm MaHPS}$ (m s$^{-1}$)  & $\mathcal{U}[-100,100]$   & $3.2^{+1.3}_{-1.3}$ & relative offset MaHPS\\
        $\mu_{\rm SOPHIE}$ (m s$^{-1}$)  & $\mathcal{U}[-35000,-34000]$ & $-34692.7^{+1.3}_{-1.3}$ & relative offset SOPHIE\\
        $\sigma_{\rm MaHPS}$ (m s$^{-1}$)  & $\mathcal{L}[0.001,100]$   & $11.5^{+1.2}_{-1.1}$ &  MaHPS jitter term\\
        $\sigma_{\rm SOPHIE}$ (m s$^{-1}$)   & $\mathcal{L}[0.001,100]$   & $6.2^{+1.2}_{-1.0}$ & SOPHIE jitter term \\
        \\
        \hline
        \\
        TOI-5108\,b Parameters \\
        \\
        $a/R_*$ & & $12.2^{+0.4}_{-0.4}$ &  Scaled semi-major axis\\
        $a$ [au] & & $0.073 \pm 0.004$ &  Semi-major axis\\
        $T_{14}$ [hours] & & $2.2 \pm 0.1$ & Transit duration of TOI-5108\,b\\
        $i$ [$^\circ$] & & $85.91 \pm 0.14$&  Orbital Inclination\\
        $R_p$ [$R_\oplus$] & & $6.6 \pm 0.1$ & Radius of TOI-5108\,b\\
        $M_p$ [$M_\oplus$] & & $32 \pm 5$&  Mass of TOI-5108\,b\\
        $\rho_p$ [kg/m$^3$] & & $600 \pm 90$&  Bulk density of TOI-5108\,b\\
        $T_{eq}$ [K] & & $1180 \pm 40$ & Equilibrium temperature of TOI-5108\,b\\
        
    \hline 
    \end{tabular}
    \begin{tablenotes}
       \item Notes: $\mathcal{U}$ indicates a uniform prior while $\mathcal{N}$ and $\mathcal{L}$ indicate normal and log-normal priors respectively. The derived values are taken from the circular model as such the eccentricity and the argument of periastron are fixed.
       \end{tablenotes}
     \end{threeparttable}
\end{table*}
\subsection{TOI-5786}
\subsubsection{Lightcurve analysis}
\begin{figure*}
    \centering
    \includegraphics[width=1.9\columnwidth]{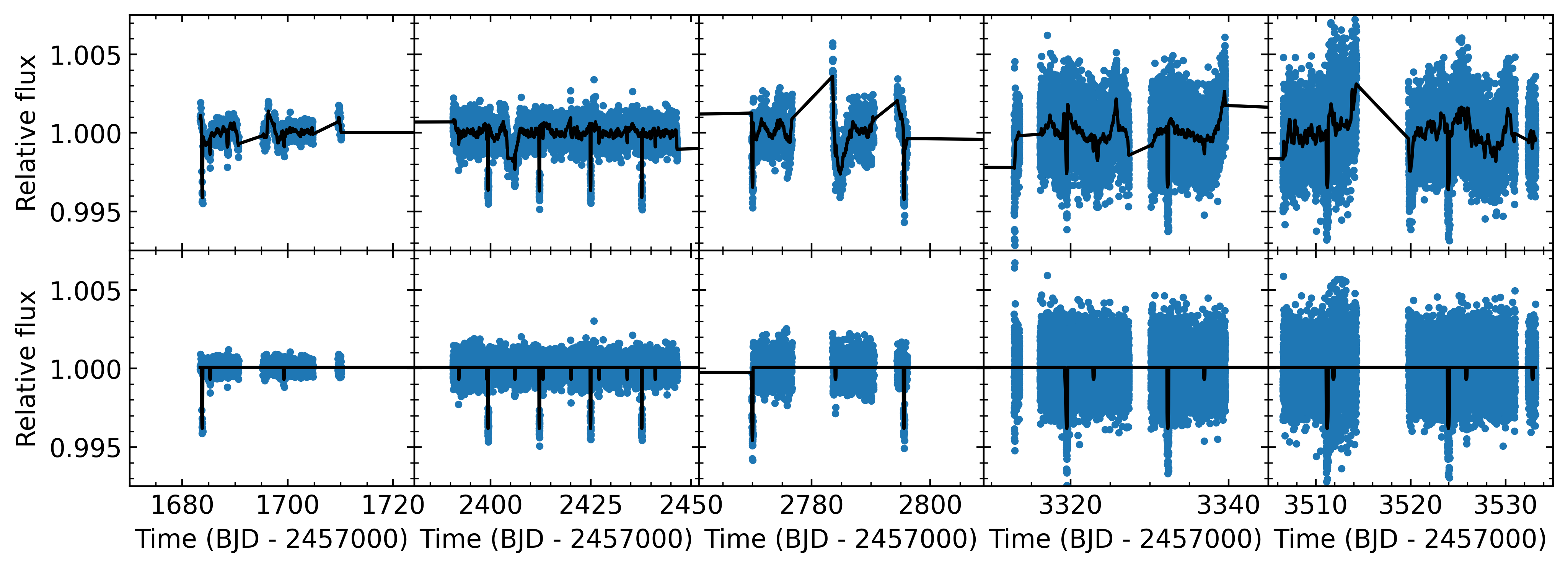} 
    \includegraphics[width=1.9\columnwidth]{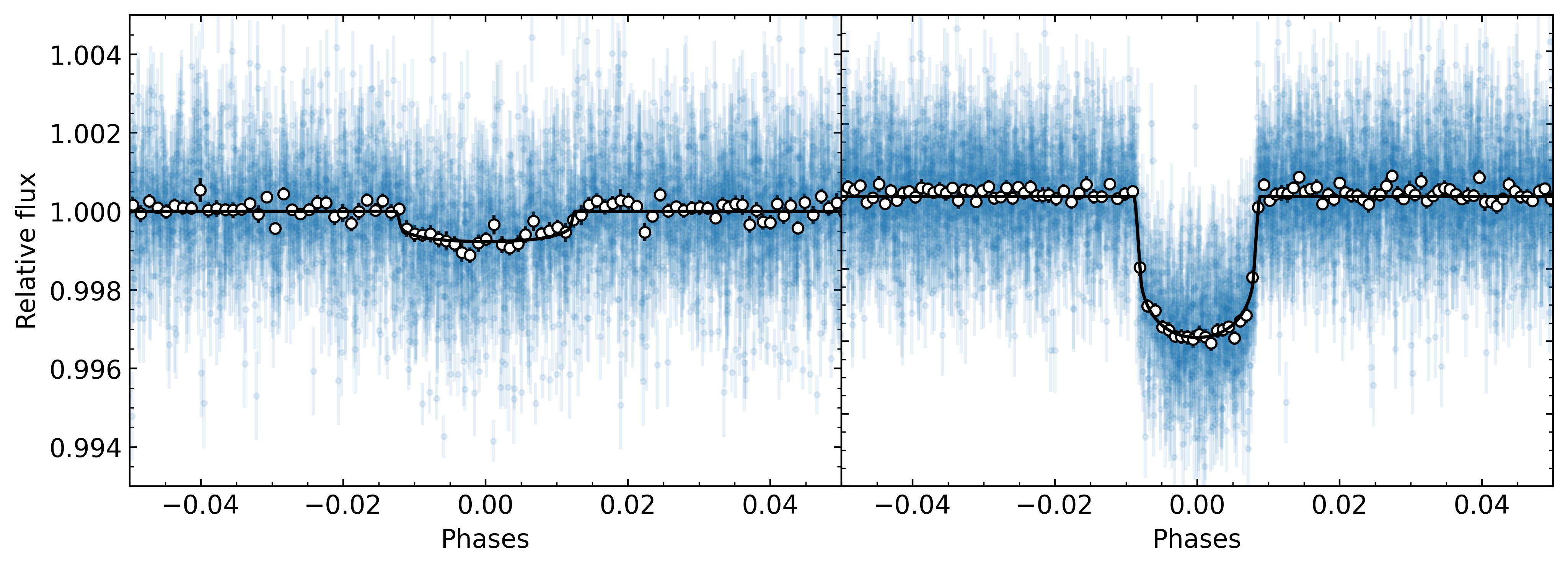} 
    \caption{Top row: \textit{TESS} lightcurve of TOI-5786 with the GP + transit model. Middle row: \textit{TESS} lightcurve with the GP model subtracted showing only the transits of the two planet candidates. Bottom row: Phase folded lightcurves of the \textit{TESS} Object of Interest TOI-5786.01 on the right and our detected planet candidate on the left.}
    \label{fig:5786fulllc}
\end{figure*}

As mentioned in Section \ref{stellarrot} the PDCSAP lightcurve for TOI-5786 shows some variation stemming from uncorrected systematics as well as potential stellar activity. As some of these variations occur close to transits, this could influence the measured transit depth and thus the derived planetary radius. To mitigate this issue we include different GP Kernels in the fitting to detrend the \textit{TESS} data. We choose three GP Kernels with varying degrees of freedom. The first Kernel is the Matern kernel parametrized in \texttt{celerite} with an amplitude ($\sigma_{GP}$) and a time/length scale ($\rho_{GP}$). This Kernel is very versatile and frequently used to model both long-term trends and short-term stochastic variations. The other two Kernels are the Simple Harmonic Oscillator (SHO) and the quasi-periodic Kernel (QP) which have both been shown to be able to reproduce the quasi-periodic oscillations associated with stellar activity quite well \citep[see e.g.][]{Ribas2018,Nicholson2022}. Details on all three Kernels are given in \cite{celerite}.
\\
In the first step, we use the \textit{TESS} light curves to fit a transit model including the known TESS candidate planet with the GP Kernels, and evaluate which configuration is preferred by our data. The Bayesian evidence values of the different fits are given in Table \ref{tab:toi5786gp}. There is a clear preference for the inclusion of a GP Kernel with all three GP fits having much higher evidence values than the fit without a GP Kernel. Among the three Kernels, the preferred one is the quasi-periodic Kernel with a strong preference ($\Delta Z > 5$) over the other two. 
\\
We also searched for additional planet candidates in the system with the Open Transit Search pipeline \citep[\href{https://github.com/hpparvi/opents}{\texttt{OpenTS;}}][]{Pope2016} using the sectors 14, 40, 41 and 54 produced by the TESS-SPOC pipeline. This analysis identified a significant transit-like signal with a period of 6.99 d that had not been reported as a TOI by the TESS mission. This planet candidate was independently detected by the Planet Hunter TESS initiative \citep{Eisner2020} and was reported as a community TOI (CTOI) on {\tt EXOFOP}.
\\
To investigate the presence of the second planet further, we fit models including the second planet to our lightcurves and compare the Bayesian evidence. We run the two-planet fit twice, once with normal priors on P and $T_0$ from the BLS analysis and once with uniform priors between 1 and 10 days for the period and 2458680 BJD and 2458690 BJD for the mid-transit time. The two-planet model is heavily favored over the one-planet model ($\Delta Z > 50$) for all cases even with uniform priors and no GP correction. Figure \ref{fig:5786fulllc} shows the \textit{TESS} lightcurve with the best fitting GP + transit model as well as the phase folded transit lightcurves of the two potential planets.
\begin{table}
    \centering
    \caption{\label{tab:toi5786gp}Bayesian evidence values for the different GP models fitted to the \textit{TESS} data of TOI-5786.}
    \begin{tabular}{lcc}
    \hline
    \hline
    Models & Evidence\\
    \hline
        No GP  & 136848.402  \\
        No GP + 2nd planet (uniform) & 136934.475 \\
        No GP + 2nd planet (normal) & 137019.481\\
        Matern & 140926.402 \\
        SHO   & 140934.706\\
        QP   &  140958.293\\
        QP + 2nd planet (uniform) & 141013.612\\
        QP + 2nd planet (normal) & 141029.223\\
    \hline 
    \end{tabular}
    
\end{table}
\subsubsection{Combined fitting}
\begin{figure*}
    \centering
    \includegraphics[width=1.9\columnwidth]{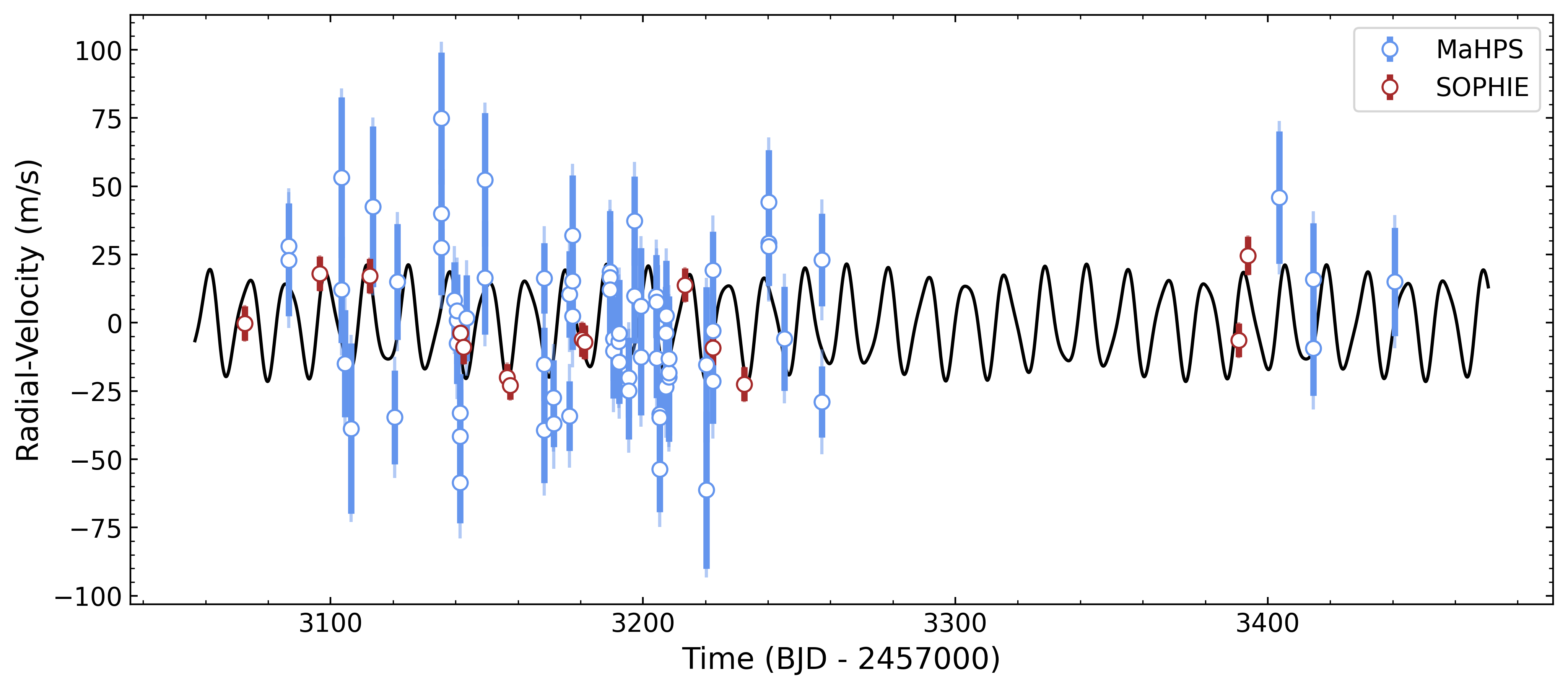} 
    \includegraphics[width=1.9\columnwidth]{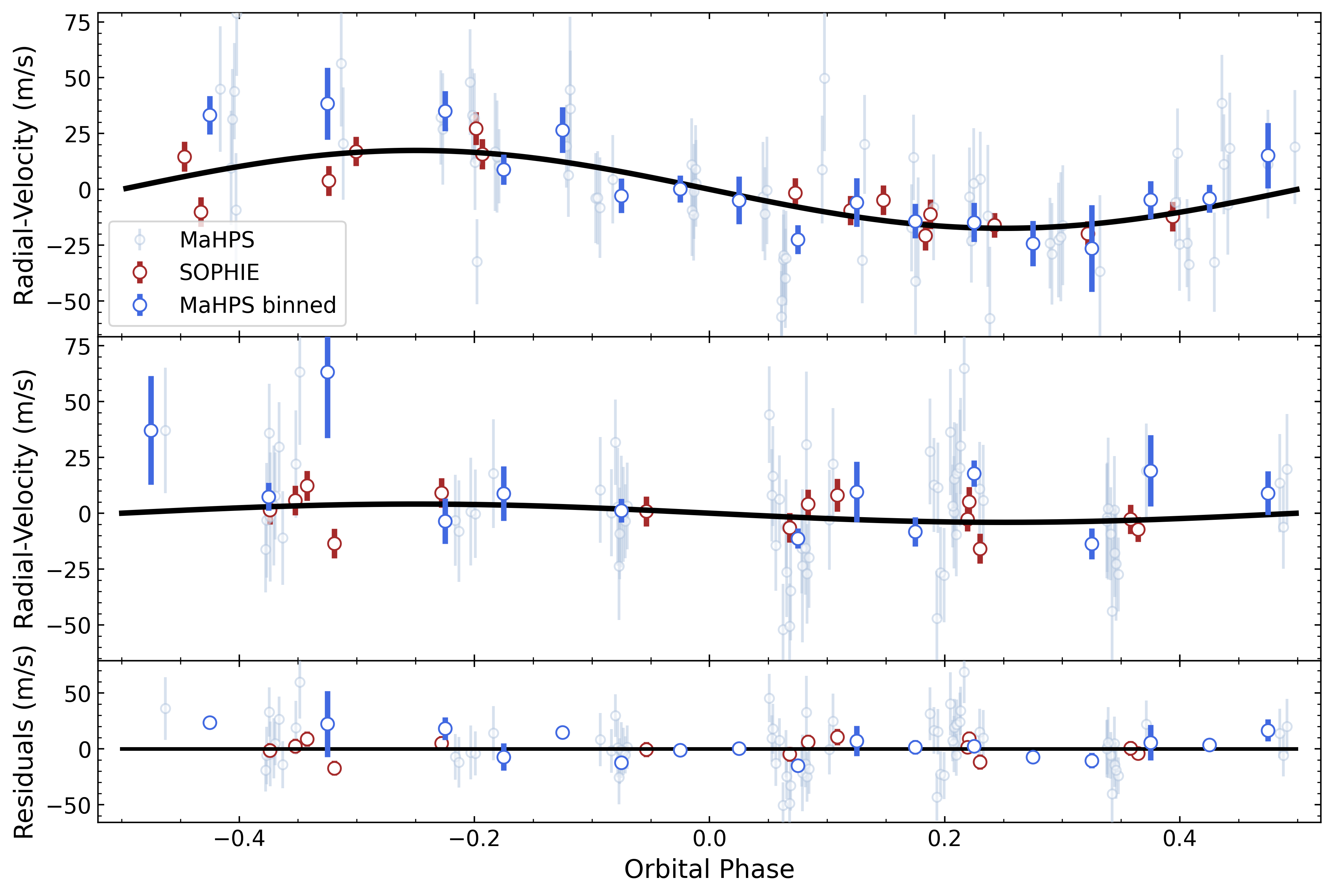} 
    \caption{First panel: Radial velocity curve of TOI-5786 with the best fit Keplerian model for the two-planet case in black. Second panel: RVs phase-folded to TOI-5786\,b. The dark blue data points represent the MaHPS RV data binned at 0.05 phases. Third Panel: RVs phase-folded to the period of the inner companion candidate. Fourth Panel: Residuals after subtracting the signals of both planets.}
    \label{fig:5786rv}
\end{figure*}

To derive the system's final parameters, we perform a joint fit of the MaHPS and SOPHIE radial velocities together with the photometry from \textit{TESS} and the two observed transits of MuSCAT2. The priors for all parameters are shown in Table \ref{tab:toi5786}. As both MuSCAT2 transit observations only included very little baseline we fixed the limb-darkening parameters to their theoretical values since it would be difficult to constrain them well from the data. For the \textit{TESS} data, we include the quasi-periodic Gaussian process kernel. We also test whether applying the QP kernel to our RV data is warranted but find such models to have a lower Bayesian evidence value compared to the ones that don't include a GP kernel on the RVs. This further supports the assumption that the GP is mainly modeling systematics in the \textit{TESS} data and not underlying stellar processes which would be present in both datasets.
\\
As with TOI-5108, the fit is performed once for a circular and once for an eccentric orbit. In our results, the circular orbit is strongly favored with a $\Delta \ln Z$ of 24.  Therefore, we adopt the values from the circular fit for the final parameters shown in Table \ref{tab:toi5786}. We are not able to measure the mass of the inner planet. However, from the fit, we derive a 3-$\sigma$ upper limit for the radial velocity amplitude of 10.0 m/s. As a consistency check, we also ran the one-planet model on the combined dataset but found that the Bayesian evidence was much lower ($\Delta \ln Z$ > 100 in favor of the two-planet model). The derived values of the orbital parameters of TOI-5786.01 are also consistent within 1$\sigma$ between the one-planet and the two-planet model. Furthermore, the RV amplitudes between a MaHPS-only and a SOPHIE-only fit are again consistent within the errors ($19 \pm 4$~m/s vs $16 \pm 3$~m/s).
\\
Figure \ref{fig:5786rv} shows the RV time series with the Keplerian model of both planets and the phase-folded RV of the outer planet.
\subsubsection{Photodynamical modeling}
We carry out a photodynamical analysis modeling the TESS photometry and the MaPHS and SOPHIE radial velocities jointly using \texttt{PyTTV} as described in \citet{2023A&A...675A.115K} to see whether we could constrain the inner planet's parameter further when accounting for gravitational interaction between the two planets. The code models the observations simultaneously using \texttt{rebound} \citep{rebound, reboundias15, reboundx} for dynamical integration and \texttt{PyTransit} \citep{Parviainen2015, Parviainen2020a, Parviainen2020b} for transit modeling, and provides posterior densities for the model parameters estimated using MCMC sampling in a standard Bayesian parameter estimation framework. The photodynamical modeling gives a smaller upper mass limit of 26.5~$M_\oplus$, and supports small eccentricities for both planets, with 95th posterior percentiles of 0.07. The analysis also predicts measurable transit timing variations (TTVs) for both planets. For the inner planet, amplitudes up to 20~min, and for the outer planet, up to 3~min are predicted. However, the sparse sampling of the transit observations combined with low signal-to-noise compared to the expected TTV amplitudes prevents a secure TTV detection in the observed photometry.
\\
\begin{table*}
    \centering
    \caption{\label{tab:toi5786}Priors and final parameter values from the \texttt{juliet} fit of TOI-5786.}
    \resizebox{\textwidth}{!}{
    \begin{tabular}{lccl}
    \hline
    \hline
    \\
    Parameter Name & Prior & Derived Values  & Description \\
    \\
    \hline
    \\
    Fitting Parameters \\
    \\
        $P_1$ [d]        &      $\mathcal{N}[6.99,0.01]$       &  $6.998405^{+0.000015}_{-0.000021}$    & Orbital Period of the inner planet\\
        $T_{0,1}$ (BJD$_{\rm TDB} -2457000$)   & $\mathcal{N}[1685.29 ,0.01]$  & $1685.288 \pm 0.001$   & Time of Mid-Transit of the inner planet\\
        $R_{p,1} / R_*$ & $\mathcal{U}[0.001, 1]$ & $0.0258 \pm 0.0011$  &  Planet-to-star radius ratio  of the inner planet\\
        $b_1$ & $\mathcal{U}[0, 1]$& $0.38^{+0.11}_{-0.14}$ &  Impact parameter  of the inner planet \\
        $\sqrt{e} \sin \omega_1$ & fixed & 0  & Parametrization of $e$ and $\omega$ \\
        $\sqrt{e} \cos \omega_1$ & fixed& 0  & Parametrization of $e$ and $\omega$\\
        $K_1$ [m~s$^{-1}$]      & $\mathcal{U}[0,200]$       &  $4.1^{+2.1}_{-1.9}$  & RV Amplitude of the inner planet\\
        $P_2$ [d]        &      $\mathcal{U}[12.7, 12.85]$       &  $12.779107 \pm 0.000015$    & Orbital Period of the outer planet\\
        $T_{0,2}$ (BJD$_{\rm TDB} -2457000$)   & $\mathcal{U}[3140.5, 3140.7]$  & $3140.6139 \pm 0.0004$   & Time of Mid-Transit of the outer planet\\
        $R_{p,2} / R_*$ & $\mathcal{U}[0.001, 1]$ & $0.0576^{+0.0009}_{-0.0008}$  &  Planet-to-star radius ratio  of the outer planet\\
        $b_2$ & $\mathcal{U}[0, 1]$& $0.33^{+0.09}_{-0.14}$ &  Impact parameter  of the outer planet \\
        $\sqrt{e} \sin \omega_2$ & fixed & 0  & Parametrization of $e$ and $\omega$ \\
        $\sqrt{e} \cos \omega_2$ & fixed& 0  & Parametrization of $e$ and $\omega$\\
        $K_2$ [m~s$^{-1}$]      & $\mathcal{U}[0,100]$       &  $17.4^{+2.1}_{-2.0}$  & RV Amplitude of the outer planet\\
        
        $\rho_*$ [kg/$\text{m}^3$]& $\mathcal{N}[690,80]$ & $682^{+90}_{-80}$ & Stellar Density\\
        \\
        Instrumental Parameters \\
        \\
        $q_{1, TESS}$ & $\mathcal{N}[0.32,0.1]$ & $0.38^{+0.07}_{-0.06}$ & limb-darkening parameter\\
        $q_{2, TESS}$ & $\mathcal{N}[0.36,0.1]$ & $0.31 \pm 0.09$ & limb-darkening parameter\\
        $\mu_{\rm MaHPS}$ (m s$^{-1}$)  & $\mathcal{U}[-100,100]$   & $1.1 \pm 2.3$ & relative offset MaHPS\\
        $\mu_{\rm SOPHIE}$ (m s$^{-1}$)  & $\mathcal{U}[-29000,-28000]$ & $-28388 \pm 2$ & relative offset SOPHIE\\
        $\sigma_{\rm MaHPS}$ (m s$^{-1}$)  & $\mathcal{L}[0.001,100]$   & $14.12 \pm 3$ &  MaHPS jitter term\\
        $\sigma_{\rm SOPHIE}$ (m s$^{-1}$)   & $\mathcal{L}[0.001,100]$   & $2.3^{+4.0}_{-2.0}$ & SOPHIE jitter term \\
        \\
        \hline
        \\
        TOI-5786\,b parameters \\
        \\
        $a/R_*$ & & $18.0 \pm 0.8$ &  Scaled semi-major axis\\
        $a$ [au] & & $0.114 \pm 0.007$ &  Semi-major axis\\
        $T_{14}$ [hours] & & $4.6 \pm 0.3$ & Transit duration of TOI-5786\,b\\
        $i$ [$^\circ$] & & $88.9 \pm 0.4$&  Orbital Inclination\\
        $R_p$ [$R_\oplus$] & & $8.54 \pm 0.13$ & Radius of TOI-5786\,b\\
        $M_p$ [$M_\oplus$] & & $73 \pm 9$&  Mass of TOI-5786\,b\\
        $\rho_p$ [kg/m$^3$] & & $640 \pm 80$&  Bulk density of TOI-5786\,b\\
        $T_{eq}$ [K] & & $1040 \pm 40$ & Equilibrium temperature of TOI-5786\,b\\
        \\
        inner planet candidate parameters \\
        \\
        $a/R_*$ & & $12.1 \pm 0.5$ &  Scaled semi-major axis\\
        $a$ [au] & & $0.077 \pm 0.004$ &  Semi-major axis\\
        $T_{14}$ [hours] & & $4.0 \pm 0.5$ & Transit duration of the inner planet candidate\\
        $i$ [$^\circ$] & & $88.2 \pm 0.7$&  Orbital Inclination\\
        $R_p$ [$R_\oplus$] & & $3.83 \pm 0.16$ & Radius of the inner planet candidate\\
        $M_p$ [$M_\oplus$] & & $<36$&  3-$\sigma$ upper limit for the mass\\
        $T_{eq}$ [K] & & $1260 \pm 40$ & Equilibrium temperature of the inner planet candidate\\
         
    \hline 
    \end{tabular}
    }
\end{table*}

\section{Results and discussion} \label{results_disc}

\subsection{Planetary parameters of the two sub-Saturns}
\begin{figure}
    \centering
    \includegraphics[width=1\columnwidth]{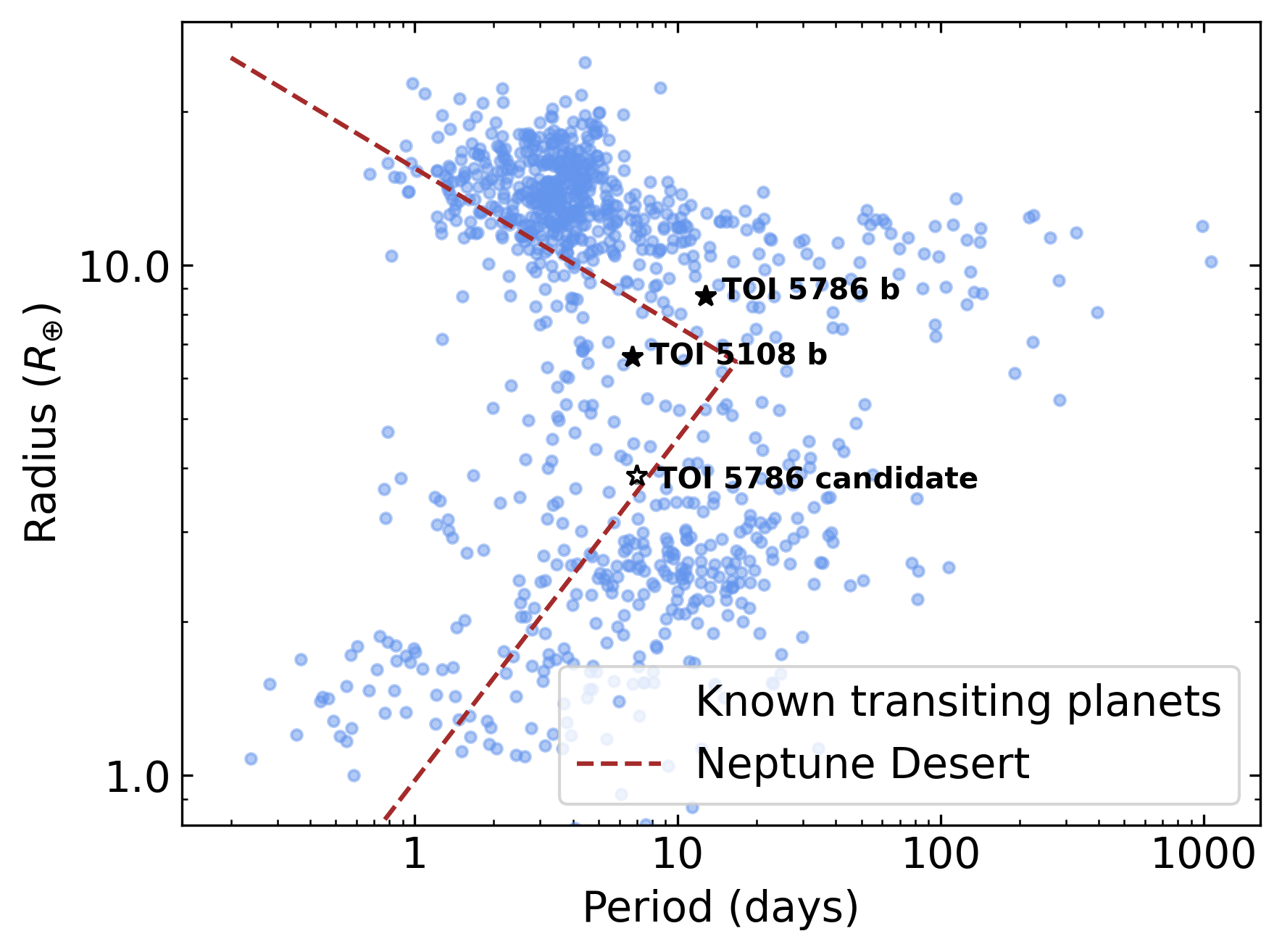} 
    \caption{The population of known transiting exoplanets with masses and radii measured to 33 \% or better taken from the NASA Exoplanet Archive (accessed on 14.10.2024). TOI-5108\,b and TOI-5786\,b are shown with black stars. The position of the planet candidate around TOI-5786 is shown with a white star. The dotted brown line indicates the boundaries of the Neptune desert as defined by \cite{Mazeh2016}.}
    \label{fig:pop}
\end{figure}

We have established the planetary nature of the two \textit{TESS} candidates which we will now refer to as TOI-5108\,b and TOI-5786\,b. Their position among the population of known transiting exoplanets is shown in Figure \ref{fig:pop}. TOI-5108\,b is a sub-Saturn sized planet with a radius of $6.6 \pm 0.1$ $R_\oplus$  and a mass of $32 \pm 5$ $M_\oplus$ orbiting its host star with a 6.753 day period. 
With the planet being this close to the host star (a = $0.073 \pm 0.004$ au) it is located inside the bounds of the Neptune desert as defined by \cite{Mazeh2016}. Depending on the XUV radiation of its host star the planet may experience significant mass loss via atmospheric escape although it is not in the very barren region of the desert ($P < 3$ days) where the escape is expected to be strongest. TOI-5786\,b is located outside the desert with a period of 12.779~days and a radius of $8.54 \pm 0.13$ $R_\oplus$. Given the mass of the host star derived in Section \ref{stellarsed} we calculate a planet mass of $73 \pm 9$ $M_\oplus$. 
\\
The newly detected planet candidate around TOI-5786 is also shown in Figure \ref{fig:pop}. With a period of 6.99 days and a radius of $3.83 \pm 0.16$ $R_\oplus$, it would be a sub-Neptune at the lower boundary of the Neptune desert. We derive 3-$\sigma$ upper limits for the mass of the planet candidate of 34 $M_\oplus$ from the Keplerian fit to the RVs and 26.5 $M_\oplus$ from the photodynamical modeling. These masses translate into a density upper limit of 3500~kg/$m^3$ indicating the presence of a volatile-rich atmosphere as expected from the larger radius.
\\
Both sub-Saturns have bulk densities similar to Saturn at $600 \pm 90$ kg/$m^3$ (TOI-5108\,b) and $640 \pm 80$ kg/$\mathrm{m}^3$ (TOI-5786\,b). This implies that a significant portion of the planet's mass is contained in a H/He envelope. 

\subsection{Envelope fractions} \label{envfrac}
\begin{figure}
    \centering
    \includegraphics[width=1\columnwidth]{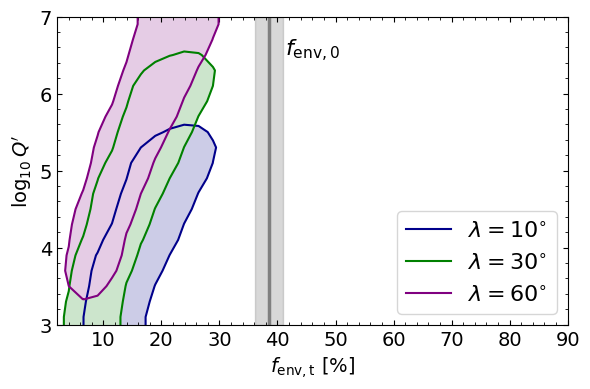} 
    \caption{Estimate of the envelope mass fraction $\text{f}_{env}$ of TOI-5108\,b using the model from \cite{Millholland2020}. The grey bar is the calculated envelope fraction with the associated error in the tides-free model. The shaded regions show the 2-$\sigma$ contours of the envelope fraction for different obliquities $\lambda$.}
    \label{fig:env}
\end{figure}

For sub-Saturns, both the rocky core and the H/He envelope contribute to the planet's mass budget. As the detailed composition of the core is assumed to only have a small effect on the overall mass fractions it is possible to constrain the envelope and core mass fractions using simple two-component models \citep{Lopez2014,Petigura2016,Petigura2017}. We use the Python code {\tt Platypos} \citep{Poppenhaeger2021} to estimate the internal structure and evolution of the two planets. The code first calculates the envelope fraction $f_{\text{env}}$ based on the models of \cite{Lopez2014}. Then it estimates the evolution of the planetary envelope given the age of the system based on energy-limited hydrodynamic atmospheric escape models \citep[see eq. 1 in ][]{Poppenhaeger2021} given the X-ray and extreme UV (XUV) irradiance of the planet. 
\\
We estimate the X-ray luminosities of our two host stars from the relations between X-ray luminosity, stellar mass, and rotation period given in \cite{Pizzolato2003}. This results in luminosities of $(1.3^{+7.9}_{-1.1}) \times 10^{28}$ erg s$^{-1}$ for TOI-5108 and $(5.5^{+32.5}_{-4.7}) \times 10^{28}$ erg s$^{-1}$ for TOI-5786. 
\\
Given this X-ray luminosity and the planetary and stellar parameters derived in this work, {\tt Platypos} estimates a present-day envelope fraction of $\sim 38 \%$ for TOI-5108\,b and $\sim 74 \%$ for TOI-5786\,b. Prevailing planet formation theory assumes that runaway accretion starts at $\text{f}_{env} \sim 50 \%$ which would mean that TOI-5786\,b was able to incite runaway accretion while the formation of TOI-5108\,b was finished shortly before. Planets with estimated envelope fraction close to $50 \%$ such as K2-19 b \citep{Petigura2020} or K2-24 c \citep{Petigura2018a} have posed a challenge to the prevailing theory of the onset of runaway accretion. 
\\
One proposed solution by \cite{Millholland2020} includes tidal heating in the model which lowers the derived envelope fractions to be more aligned with expectations from theory. Based on our analysis TOI-5108\,b does not appear to have any significant eccentricity in its orbit, however, the obliquity of the planet is unknown and a non-zero axial tilt can also cause tidal heating. We therefore use the model from \cite{Millholland2020} to calculate the envelope fraction of TOI-5108\,b including tidal heating\footnote{TOI-5786\,b is outside the parameter space of the \cite{Millholland2020} model (1 - 70 $M_\oplus$) and is therefore not included in this analysis.}. The strength of the reduction of the envelope fraction after the inclusion of tidal forces depends on the obliquity of the orbit and the tidal quality factor $Q^{\prime}$. Based on the solar system and exoplanets with measured tidal quality factors $Q^{\prime} \sim 10^4 - 10^5$ is a realistic estimate for the parameter \citep{Murray1999,Zhang2008,Morley2017,Puranam2018}. In Figure \ref{fig:env} we show the results of the derived envelope fraction for TOI-5108\,b as a function of $Q^{\prime}$ for different obliquities. Without the inclusion of tides, the envelope fraction is similar to the one calculated with {\tt Platypos} at $38.6 \pm 2.4 \ \%$. With tides, this drops to between $10 - 20 \ \%$ depending on the obliquity $\lambda$ and $Q^{\prime}$. At these envelope fractions, the internal and atmospheric structure of TOI-5108\,b would likely be more similar to a sub-Neptune than a gas giant.
\\
The expected amplitude of the Rossiter McLaughlin effect for TOI-5108 is $\sim 5$~m/s. As such it will be possible to measure the obliquity of TOI-5108~b with existing high-resolution spectrographs although a larger mirror size than the size of the telescopes used in this paper is preferable to get a better time sampling during the transit. 
\subsection{Future characterization possibilities}
\begin{figure}
    \centering
    \includegraphics[width=1\columnwidth]{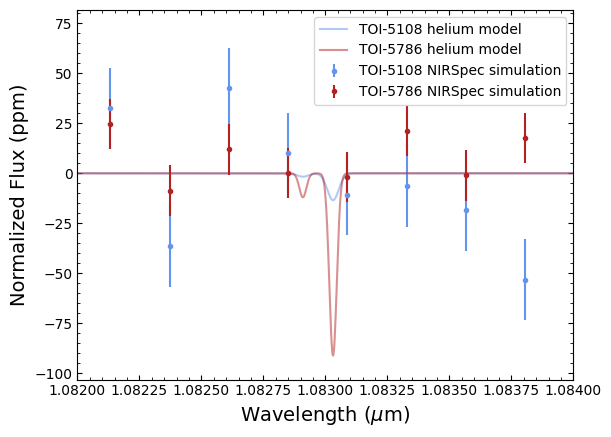} 
    \caption{Simulated transit observation with NIRSpec for the helium triplet absorption signals calculated with {\tt p-winds}. The helium absorption models shown are for the high XUV input spectrum (WASP-17) with the largest mass-loss rates.}
    \label{fig:heliumsim}
\end{figure}

Using equation 1 from \cite{Kempton2018} we calculate the Transmission Spectroscopy Metric (TSM) of both planets to assess whether they are favorable targets for atmospheric characterization with JWST. The TSM is a measure for the expected S/N of a JWST transit observation given the properties of the host star and the expected strength of spectral features in the planet's atmosphere. 
\\
We calculate TSM values of $138 \pm 25$ and $78 \pm 12$ for TOI-5108\,b and TOI-5786\,b respectively. For sub-Jovian planets, \cite{Kempton2018} found a threshold of 90 for a planet to be considered a high-quality follow-up target for JWST making TOI-5108\,b a good target to further investigate the atmospheric composition of these intermediate-sized planets. While TOI-5786\,b is slightly below this threshold, among the sub-Saturns with longer periods ($P < 10$ days) and precise masses and radii, it is in the top $20 \%$ of the highest TSM making it a solid target for further study in the slightly longer period regime.
\\
Due to its position in the Neptune desert, one point of interest for an atmospheric follow-up of TOI-5108\,b would be to search for signs of an escaping atmosphere. Atmospheric escape through photoevaporation is one of the proposed mechanisms that could result in a dearth of short-period, intermediate-sized planets. As TOI-5786\,b is located a bit farther away from its host star and also slightly more massive, the mass loss through atmospheric escape is expected to be slightly weaker. 
\\
Using the open source code {\tt p-winds} \citep{DosSantos2022}, we investigate whether ongoing atmospheric escape would be detectable through transmission spectroscopy using the metastable triplet Helium line as a tracer. Based on the planetary parameters and an XUV spectrum of the host star, {\tt p-winds} calculates the ionization fraction of hydrogen and helium and then models the atmospheric escape with a one-dimensional Parker wind model \citep{Parker1958,Oklop2018}. It then estimates the strength of the absorption from the metastable triplet helium at 1083 nm during transit.
\\
Since the XUV spectra of TOI-5108 and TOI-5786 are unknown, we use measured spectra from similar stars as proxies. To cover a range of XUV luminosities we use HD 108147, the sun, and WASP 17 as proxies of low, medium, and higher stellar activity respectively. The spectra of HD 108147 and WASP 17 are taken from the MUSCELS survey \citep{Loyd2016,Youngblood2016,France2016,Behr2023}. Due to the older age, TOI-5108\,b is likely more similar to the Sun or HD 108147 while the younger age and shorter rotation period point to TOI-5786 having a higher activity. 
\\
We first estimate the mass-loss rate of both planets based on the integrated XUV flux of our three proxy spectra with a simple energy-limited hydrodynamic escape model \citep{watson1981dynamics,lopez2012thermal,owen2012planetary}, following the assumption of \cite{foster2022exoplanet} who calculated atmospheric escape rates from eROSITA data.
The mass-loss rate is given by:
\begin{align}
\dot{M} = \epsilon \frac{\pi {R^2_{XUV}} F_{XUV}}{K_{tide} G M_{pl}/R_{pl}}
\end{align}

where \(\epsilon\) is an efficiency parameter set to 0.15, G is the gravitational constant, \(R_p\) and \(M_p\) are the planetary radius and mass, \(F_{XUV}\) is the X-ray and extreme-UV flux incident on the planet and \(R_{XUV}\) is the planetary radius at XUV wavelengths (assumed to be 1.1 times the optical radius).
\(K_{tide}\) is the reduction factor of the planetary gravitational potential due to the effect of a stellar tide, which we assume to be negligible.
\\
Table \ref{tab:fxuv} shows the integrated XUV flux and corresponding mass-loss rates of both planets for the three XUV spectra. 
\begin{table}
    \centering
    \caption{\label{tab:fxuv}X-ray and EUV fluxes and the corresponding mass-loss rates derived from the three proxy spectra for both planets.}
    \begin{tabular}{lcc}
    \hline
    \hline
    Spectrum & $F_{XUV}$ [W/m$^2$] &$\dot{M}$ [g/s] \\
    \hline
    \\
    TOI-5108\,b & & \\
    HD 108147 & 0.297  & $(1.00 \pm 0.19) \times 10^9$ \\
    Sun  & 0.560  & $(1.89 \pm 0.35) \times 10^9$ \\
    WASP 17 & 2.891 & $(9.8 \pm 1.8) \times 10^9 $ \\
    \\
    TOI-5786\,b & & \\
    HD 108147 & 0.114 & $(3.6 \pm 0.4) \times 10^8$ \\
    Sun   & 0.216 & $(6.8 \pm 0.8) \times 10^8$ \\
    WASP 17 & 1.114 & $(3.5 \pm 0.4) \times 10^9$ \\
    \hline 
    \end{tabular}
\end{table}
We use the estimated mass-loss rates, XUV spectra, and the planetary and stellar parameters as input for {\tt p-winds} to calculate the expected absorption in the triplet helium during transit assuming a hydrogen fraction of 0.9 and a thermospheric temperature of 7000 K. Next we simulate an observation with JWST NIRSpec G140H ($R \sim 2700$) using PandExo \citep{Batalha2017} to check whether the signal would be observable. As shown in Figure \ref{fig:heliumsim}, even in the case of the strongest XUV flux, the helium signal is not detectable in the NIRSpec data. Due to the lower mass-loss rates and the suboptimal stellar types with regards to the triplet helium feature \citep[see ][]{oklopcic2019} the absorption signal is too weak to be detectable. 
\\
Lastly, with the mass-loss rates from Table \ref{tab:fxuv} and the envelope fractions of Section \ref{envfrac} we can calculate the time it would take for the planets to lose their atmospheres. This yields $288 \pm 50$ Gyrs for TOI-5786\,b and $24 \pm 6$ Gyrs for TOI-5108\,b in the highest mass-loss scenario showing that both planets are presently stable against photoevaporation.

\subsection{False positive tests for the second transit signal around TOI-5786}
From our RV data, we were not able to detect the potential inner planet around TOI-5786 as the errors on the SOPHIE and especially the MaHPS data much are larger than the expected RV amplitude from the inner planet ($\sim 3$ m/s).
\\
To further validate the signal as a planet, we perform several false-positive tests on the TESS data. First, we split the transit dataset into odd and even transits to investigate whether there is a difference in transit depth indicative of an eclipsing binary configuration. We find transit depths of $1063 \pm 60$~ppm for the odd transits and $1058 \pm 90$~ppm for the even transits. These depths agree well within the uncertainties indicating a consistent transit depth between the odd and even transits.
\\
Next, we investigate the source of the transit event by looking at the difference images between the out-of-transit and the in-transit flux created from the target pixel files that are created by the SPOC. Due to the low signal-to-noise of the individual transits the pixel with the highest flux difference is not always associated with TOI-5786. For multiple transits, the pixel with the highest difference in flux before and during the transit is close to TIC 40292668. However, a visual inspection of the lightcurve of TIC 40292668 shows no transit events that coincide with the transit signal of the potential inner planet of TOI-5786. Furthermore, the lightcurve of TIC 40292668 shows some variability which could explain the offset of flux difference images. 
\\
Lastly, we use the statistical validation tool TRICERATOPS \citep{Giacalone2020,Giacalone2021} to estimate the false-positive probability of our transit signal. TRICERATOPS is a Bayesian tool that compares the derived lightcurve to a wide range of possible transit-producing events, taking into account the used aperture as well as knowledge of the target star and its surrounding stars from GAIA to calculate a false positive probability (FPP) and a nearby false positive probability (NFPP). We use data from all available sectors for TOI-5786 to produce a phase-folded lightcurve which we use as an input for the analysis. First, we calculate the SNR of our transits using Equation 17 of \cite{Giacalone2021} to assess whether the derived FPP and NFPP from Triceratops are reliable. We get an SNR of 30 which is above the threshold of 15 for reliable results as found in \cite{Giacalone2021}. We run TRICERATOPS 20 times and derive an average FPP of $0.030 \pm 0.004$ and a NFPP of $0.0154 \pm 0.0014$. The bulk of the NFPP probability is caused by TIC 40292668 (an average of 0.14) which does not show any obvious transit events in its lightcurve. While these probabilities are rather low, they are not enough to statistically validate the planet according to \cite{Giacalone2021} (FPP < 0.015 and NFPP < $10^{-3}$). 
\\
Since we detected a secondary star close to TOI-5786 through our high-resolution imaging observations we consider the scenario where the second planet is orbiting this star. Given the magnitude difference between the two stars ($\sim 5.3$~mag) in the $I_\mathrm{c}$ filter and the measured transit depth on TOI-5786 of 690~ppm, the depth on the secondary star would need to be $\sim 9~\%$ making it most likely a giant planet or brown dwarf transiting the M star. The TRICERATOPS probability for a blended eclipsing binary or planet transit on a blended star is rather low ($0.015 \pm 0.004$ adding the different blended scenarios). To further investigate this possibility we fit the TOI-5786 lightcurve with single-planet models for TOI-5786 b and for the second planet candidate. In this fit, we impose a uniform prior on the stellar density. We can then compare the derived densities to the stellar density from Section \ref{stellar} to asses which star is more likely to host each planet \citep[see][]{Lester2022}. The derived density for the transit fit of TOI-5786~b is $789^{+50}_{-120}$~kg/$m^3$  which is similar to the depth derived from the transit fit of the second planet candidate at $713^{+170}_{-360}$~kg/$m^3$. Both densities agree well with the value derived in Section \ref{stellar} of $690 \pm 80$~kg/$m^3$. We therefore conclude that it is unlikely, that the second transiting signal happens on the blended stellar companion.

\section{Conclusions and summary}
We have detected and measured precise masses for the new transiting planets TOI-5108\,b and TOI-5786\,b. The planets are sub-Saturns with radii of $6.6 \pm 0.1$~$R_\oplus$ and $8.54 \pm 0.13$~$R_\oplus$, and masses of $32 \pm 5$~$M_\oplus$ and $73 \pm 9$~$M_\oplus$ respectively. Additionally, we find a signal of another transiting planet in the TESS lightcurve of TOI-5786. This potential planet would be an inner companion to TOI-5786\,b with a period of 6.99 days and a radius of $3.83 \pm 0.16$ $R_\oplus$. With a period of 6.573 days, TOI-5108\,b is located inside the bounds of the Neptune desert whereas TOI-5786\,b with a period of 12.779~days is located at the upper edge.
\\
The metallicity of TOI-5108 ($0.20^{+0.02}_{-0.05}$ dex) suggests that the envelope of TOI-5108\,b could be metal-rich which would have helped the planet retain a large fraction of its atmosphere against photoevaporation during its early lifetime due to the enhanced cooling from atomic metal lines. TOI-5108\,b follows the trend of hot super-Neptune-sized planets preferably occurring around metal-rich stars. Even TOI-5786 has a slightly higher metallicity although it is consistent with solar metallicity within the uncertainty.
\\
TOI-5108\,b and TOI-5786\,b have envelope mass fractions slightly below and above the runaway accretion limit at $ \sim 38 \%$ and $ \sim 74 \%$ respectively. As both planets have high TSM values, an in-depth study of their respective atmospheres could be used to look for differences in the composition of planets that were able to incite runaway gas accretion and planets that stopped forming earlier. 
\\
From our analysis, we find that both planets are likely stable against photoevaporation. However, our calculation of the mass-loss rates is based on proxy spectra, so a direct measurement of the mass-loss rates through transmission spectroscopy is needed to draw definitive conclusions. Even if the mass-loss signature is not detectable as predicted from our simulations, an upper limit on the mass-loss rate could be used to confirm the assumption of stability against photoevaporation. 
\\
The stability of the two planets against photoevaporation further strengthens the assumption that the upper edge of the Neptune desert is caused by a different mechanism such as high-eccentricity migration. Interestingly, an inner companion for TOI-5786\,b would reduce the likeliness of high-eccentricity migration as the mechanism for its present-day orbit. Such a planetary system would likely form through a more quiet mechanism like disk migration. This does not necessarily conflict with the assumption that the upper edge of the Neptune desert is formed by high-eccentricity migration because TOI-5786\,b is not located directly at the desert boundary.

\begin{acknowledgements}
We thank the anonymous referee for the corrections and suggestions that helped us improve the presentation of the paper.
\\
We want to acknowledge the independent detection of the inner planet around TOI-5786 by the Planet Hunter TESS team including Sam Lee, Hans Martin Schwengeler, Cledison Marcos da Silva, Simon Lund Sig Bentzen, Kim Donghyeon, Safaa Alhassan, Ciaron Drain, Marc Huten, Roberto Gorelli, Allan Tarr, Ian R. Mason and Nora Eisner. We note that they also found two more potential transit events that are not associated with the two planets. However, as these events are separated by 1634 days and the period cannot be determined they are not considered in our analysis.
\\
LT acknowledges support from the Excellence Cluster ORIGINS funded by the Deutsche Forschungsgemeinschaft (DFG, German Research Foundation) under Germany's Excellence Strategy – EXC 2094 – 390783311.
\\
The Wendelstein 2.1 m telescope project was funded by the Bavarian government and by the German Federal government through a common funding process. Part of the 2.1 m instrumentation including some of the upgrades for the infrastructure were funded by the Cluster of Excellence “Origin of the Universe” of the German Science foundation DFG.
\\
We thank the Observatoire de Haute-Provence (CNRS) staff for its support. This work was supported by the ``Programme National de Plan\'etologie'' (PNP) of CNRS/INSU and CNES.
\\
Funding for the \textit{TESS} mission is provided by NASA's Science Mission Directorate. KAC and CNW acknowledge support from the \textit{TESS} mission via subaward s3449 from MIT.
\\
We acknowledge the use of public \textit{TESS} data from pipelines at the TESS Science Office and at the TESS Science Processing Operations Center. 
\\
Resources supporting this work were provided by the NASA High-End Computing (HEC) Program through the NASA Advanced Supercomputing (NAS) Division at Ames Research Center for the production of the SPOC data products.
\\
This paper includes data collected by the \textit{TESS} mission, which are publicly available from the Mikulski Archive for Space Telescopes (MAST) operated by the Space Telescope Science Institute (STScI).
\\
This research has made use of the Exoplanet Follow-up Observation Program (ExoFOP; DOI: 10.26134/ExoFOP5) website, which is operated by the California Institute of Technology, under contract with the National Aeronautics and Space Administration under the Exoplanet Exploration Program.
\\
This work makes use of observations from the LCOGT network. Part of the LCOGT telescope time was granted by NOIRLab through the Mid-Scale Innovations Program (MSIP). MSIP is funded by NSF.
\\
We acknowledge financial support from the Agencia Estatal de Investigaci\'on of the Ministerio de Ciencia e Innovaci\'on MCIN/AEI/10.13039/501100011033 and the ERDF “A way of making Europe” through project PID2021-125627OB-C32, and from the Centre of Excellence “Severo Ochoa” award to the Instituto de Astrofisica de Canarias.
\\
This article is based on observations made with the MuSCAT2 instrument, developed by ABC, at Telescopio Carlos S\'{a}nchez operated on the island of Tenerife by the IAC in the Spanish Observatorio del Teide. This work is partly financed by the Spanish Ministry of Economics and Competitiveness through grants PGC2018-098153-B-C31.
\\
This work is partly supported by JSPS KAKENHI Grant Numbers JP24H00017, JP24H00248, JP24K00689, JP24K17082 and JSPS Bilateral Program Number JPJSBP120249910.
\\
This research has made use of the NASA Exoplanet Archive, which is operated by the California Institute of Technology, under contract with the National Aeronautics and Space Administration under the Exoplanet Exploration Program.
\\
S.G.S acknowledges the support from FCT through Investigador FCT contract nr. CEECIND/00826/2018 and  POPH/FSE (EC).
A.A.B. and I.A.S. acknowledge the support of M.V. Lomonosov Moscow State University Program of Development.
DRC acknowledges partial support from NASA Grant 18-2XRP18\_2-0007. Based on observations obtained at the Hale Telescope, Palomar Observatory, as part of a collaborative agreement between the Caltech Optical Observatories and the Jet Propulsion Laboratory(operated by Caltech for NASA). 
This work was supported by a NASA Keck PI Data Award, administered by the NASA Exoplanet Science Institute. Data presented herein were obtained at the W. M. Keck Observatory from telescope time allocated to the National Aeronautics and Space Administration through the agency's scientific partnership with the California Institute of Technology and the University of California. 
The Observatory was made possible by the generous financial support of the W. M. Keck Foundation. 
The authors wish to recognize and acknowledge the very significant cultural role and reverence that the summit of Maunakea has always had within the indigenous Hawaiian community. We are most fortunate to have the opportunity to conduct observations from this mountain.
This research was carried out in part at the Jet Propulsion Laboratory, California Institute of Technology, under a contract with the National Aeronautics and Space Administration (80NM0018D0004).
\end{acknowledgements}

\bibliographystyle{aa}
\bibliography{bib} 
\begin{appendix}
\section{Transit lightcurves}

\begin{minipage}{\textwidth}
    \centering
    \includegraphics[width=0.45\textwidth]{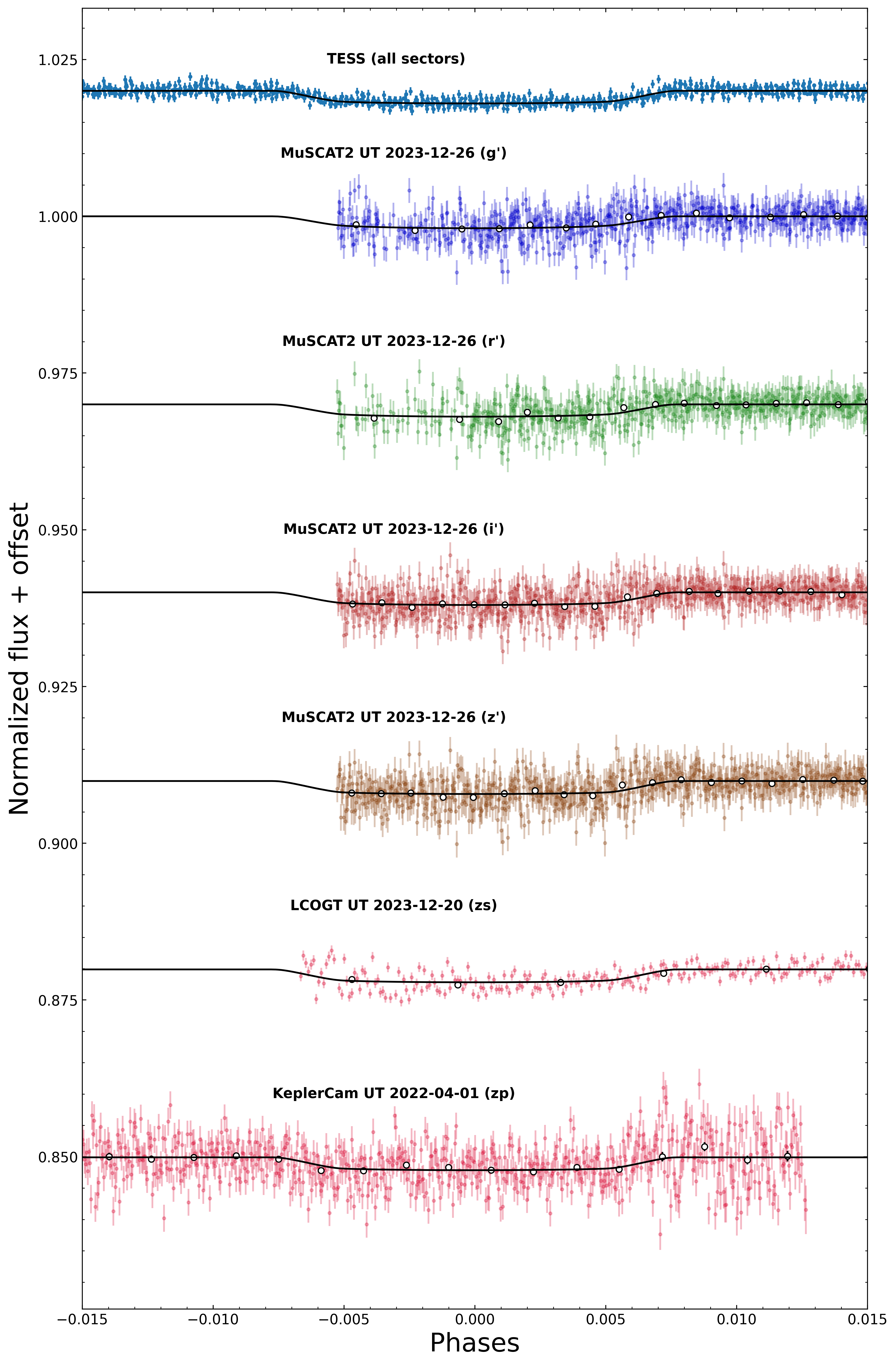}
    \includegraphics[width=0.45\textwidth]{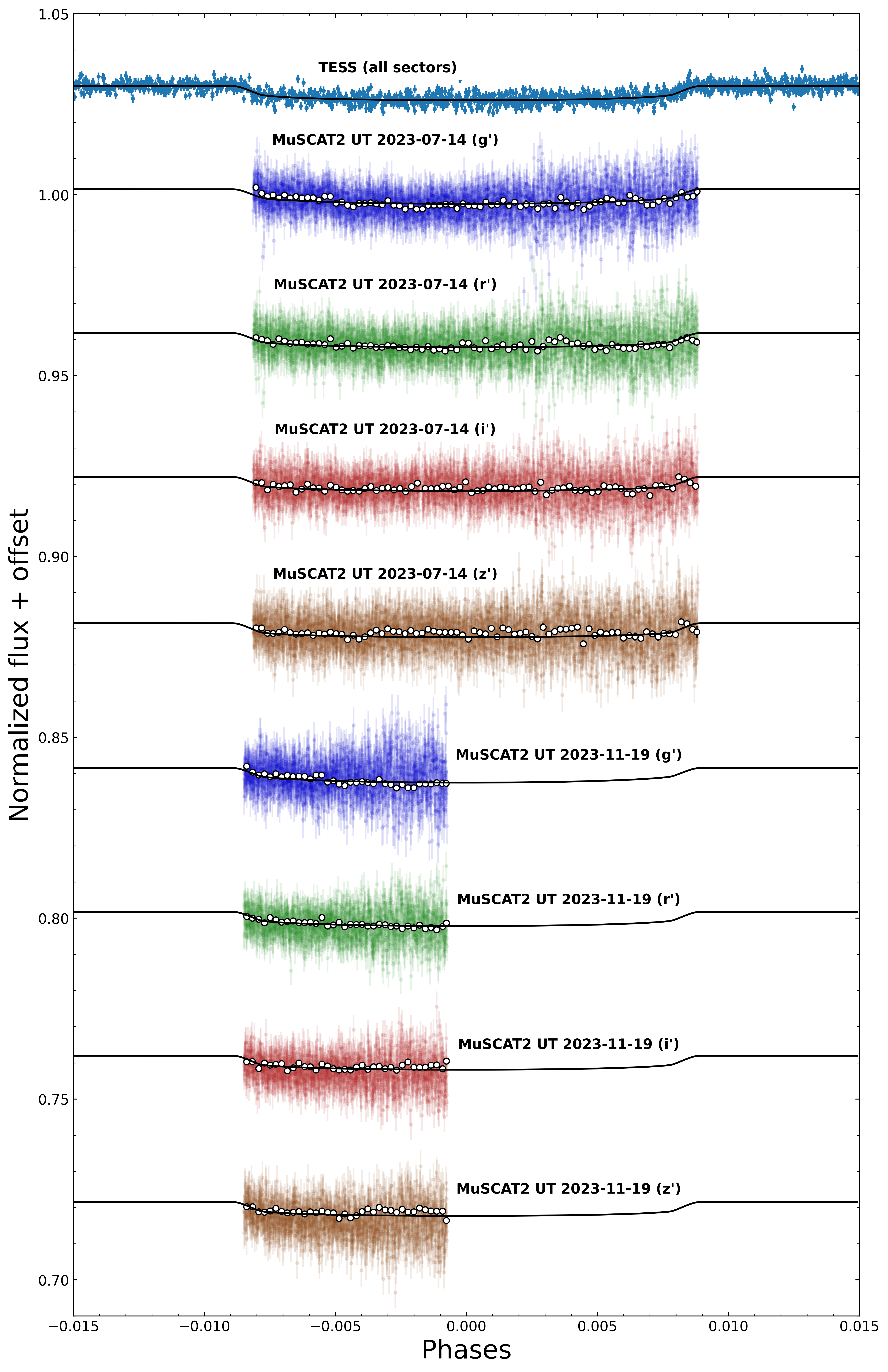}
    \captionof{figure}{Transit lightcurves of TOI-5108 (left) and TOI-5786 (right). The \textit{TESS} lightcurves are phase-folded with the data from all available sectors.}
    \label{fig:lcall}
\end{minipage}
\FloatBarrier
\section{Stellar parameter comparison}
\begin{minipage}{\textwidth}
    \centering
    \captionof{table}{Comparison of the derived stellar parameters in Section \ref{stellarsed} using SED and isochrone fitting.}
    \resizebox{\textwidth}{!}{
    \begin{tabular}{lcccccc}
    \hline
    \hline
    & &  TOI-5108 && &TOI-5786&\\
      Parameters   & SED + iso & {\tt isoclassify} (MIST) & {\tt isoclassify} (PARSEC)  & SED + iso & {\tt isoclassify} (MIST) & {\tt isoclassify} (PARSEC) \\
      \hline
        \\
        T$_{\text{eff}}$ [K]& $5818 \pm 30$ & $5808 \pm 40$ & $5818^{+20}_{-50}$& $6240 \pm 40$ & $6235 \pm 50$ &$6237^{+50}_{-40}$\\
        $\log g$ [cgs]& $4.18 \pm 0.06$ & $4.26^{+0.05}_{-0.04}$ & $4.25 \pm 0.03$ & $4.22^{+0.06}_{-0.07}$ & $4.26^{+0.04}_{-0.03}$ &$4.26 \pm 0.03$\\ 
        $\left[\textrm{Fe/H}\right]$ [dex] & $0.20 \pm 0.04$ & $0.20^{+0.02}_{-0.05}$ & $0.19^{+0.02}_{-0.05}$  & $0.11 \pm 0.04$ & $0.08^{+0.1}_{-0.08}$& $0.12^{+0.06}_{-0.13}$\\
        R$_*$ [R$_\odot$] &  $1.32 \pm 0.02$ & $1.29 \pm 0.04$ & $1.29 \pm 0.04$  & $1.36 \pm 0.02$& $1.36^{+0.03}_{-0.05}$ & $1.35 \pm 0.03$ \\ 
        M$_*$ [M$_\odot$] & $1.14^{+0.02}_{-0.08}$ & $1.10 \pm 0.02$ & $1.08^{+0.02}_{-0.03}$  & $1.29^{+0.03}_{-0.06}$ & $1.23^{+0.04}_{-0.03}$& $1.25^{+0.01}_{-0.06}$\\ 
       L$_{bol}$ [L$_\odot$] & $1.80 \pm 0.05$ & $1.71^{+0.09}_{-0.08}$ & $1.70 \pm 0.08$ & $2.52 \pm 0.08$& $2.52^{+0.08}_{-0.12}$  & $2.51^{+0.09}_{-0.10}$\\
       Age [Gyr] & $5.2^{+2.2}_{-0.5}$ & $5.8^{+1.0}_{-0.8}$ & $6.3^{+1.0}_{-0.9}$  & $2.2^{+0.2}_{-1.0}$ & $2.5^{+0.5}_{-0.7}$& $2.6^{+0.7}_{-0.5}$ \\ 
       D [pc] & $130.8^{+0.5}_{-0.3}$ & $131.1 \pm 0.06$ & $131.1 \pm 0.06$ & $193.6^{+0.5}_{-0.4}$ & $193.8 \pm 0.05$& $193.8 \pm 0.05$ \\
       A$_V$ [mag]& $0.02 \pm 0.02$ & - & - & $0.06^{+0.04}_{-0.03}$ & - & -\\ 
       \hline
    \end{tabular}
    \label{tab:starcomp}
    }
\end{minipage}
\onecolumn
\section{High-resolution imaging}
\begin{minipage}{\textwidth}
    \centering
    \includegraphics[width=0.45\textwidth]{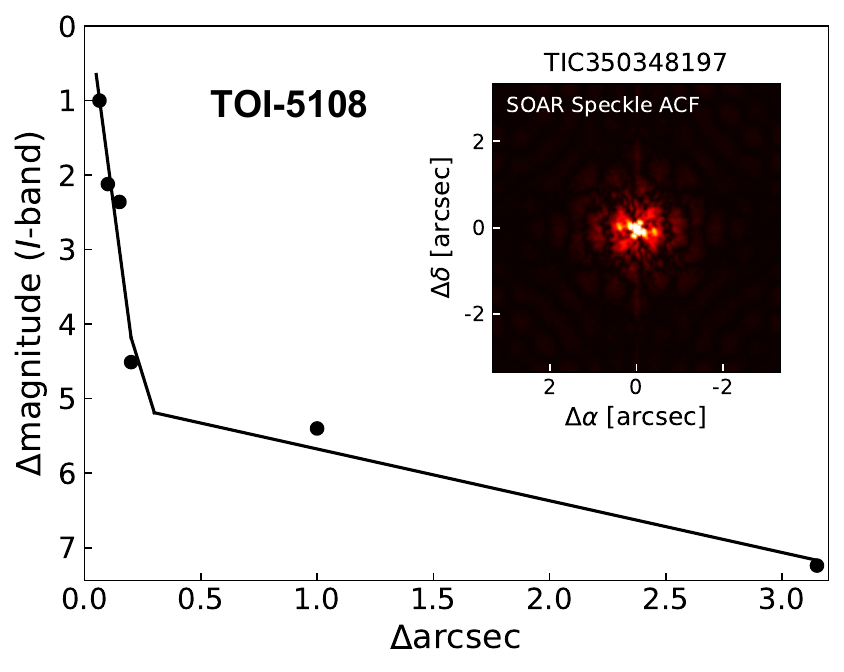} 
    \includegraphics[width=0.49\textwidth]{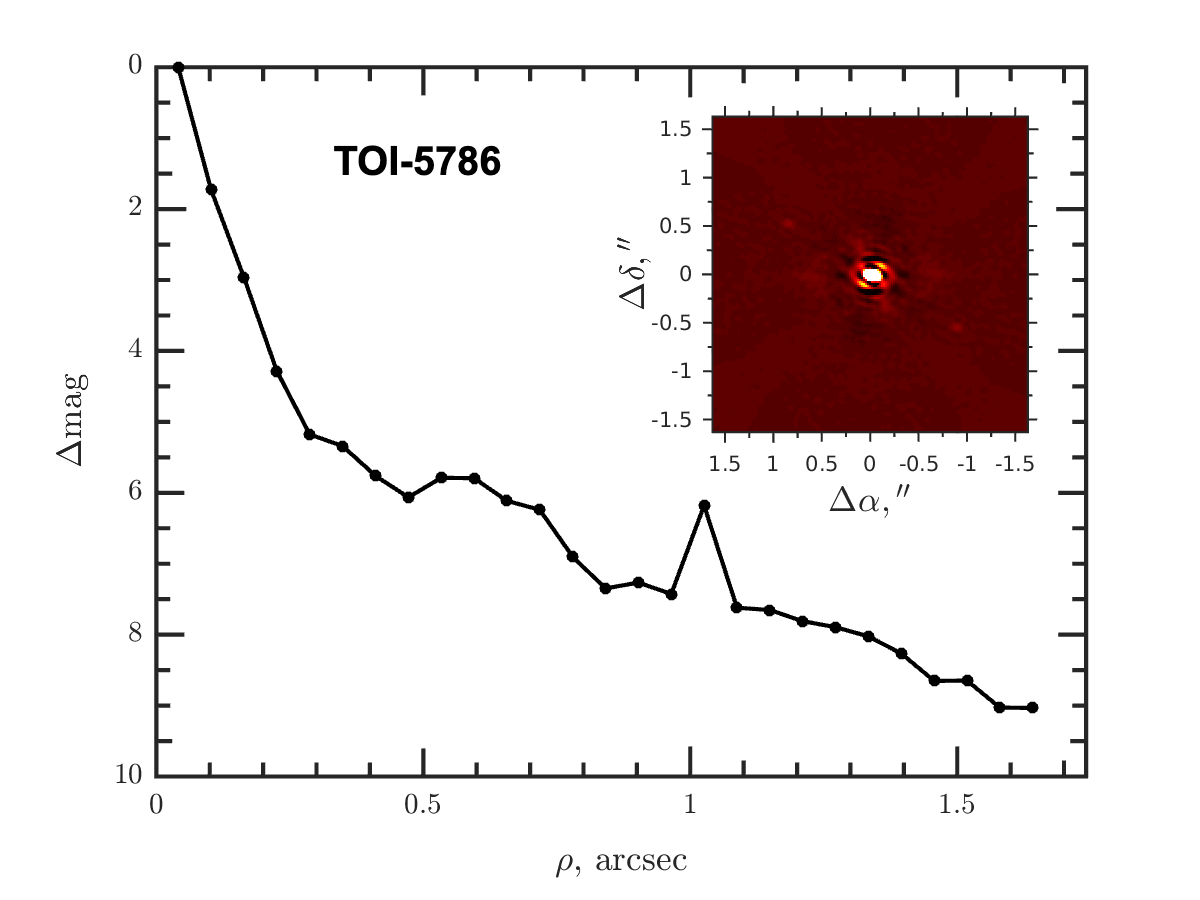} 
    \includegraphics[width=0.49\textwidth]{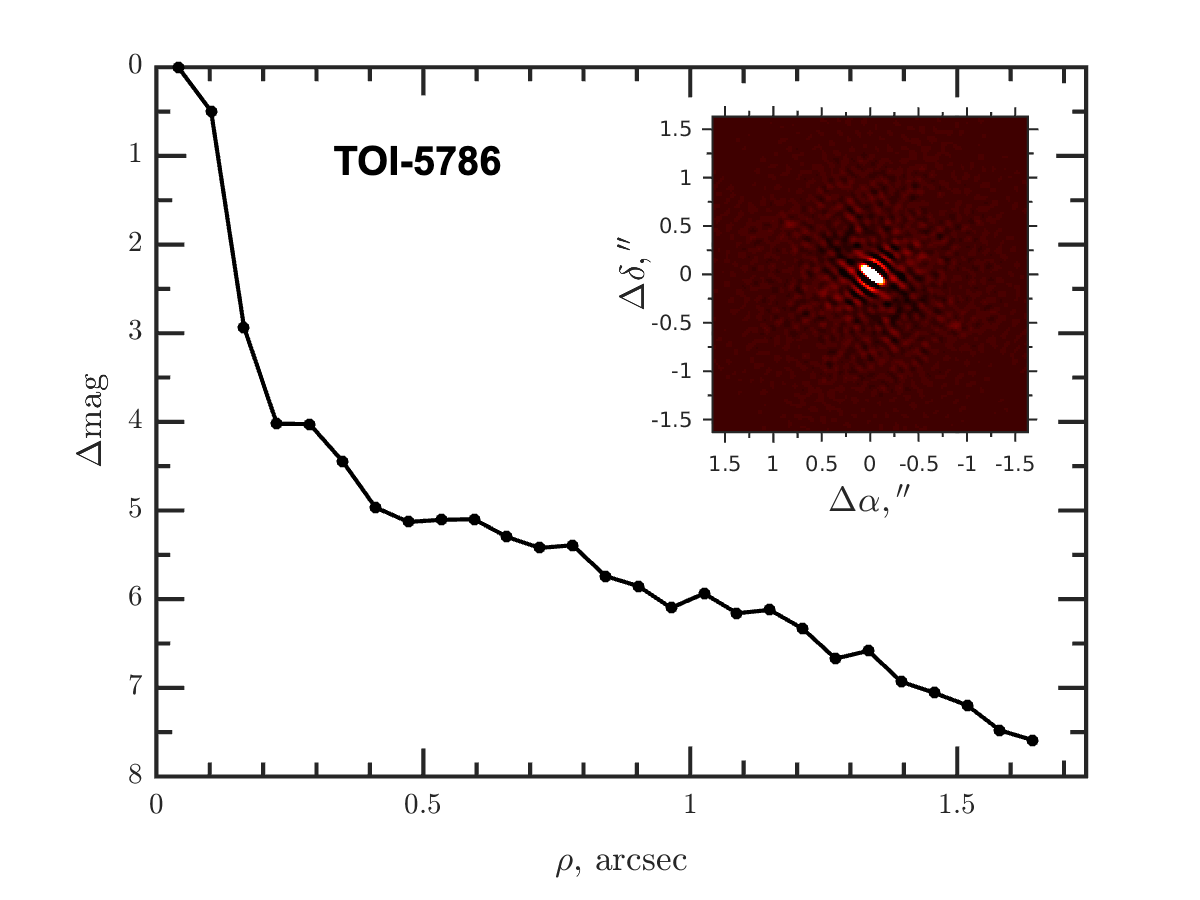}
    \includegraphics[width=0.49\textwidth]{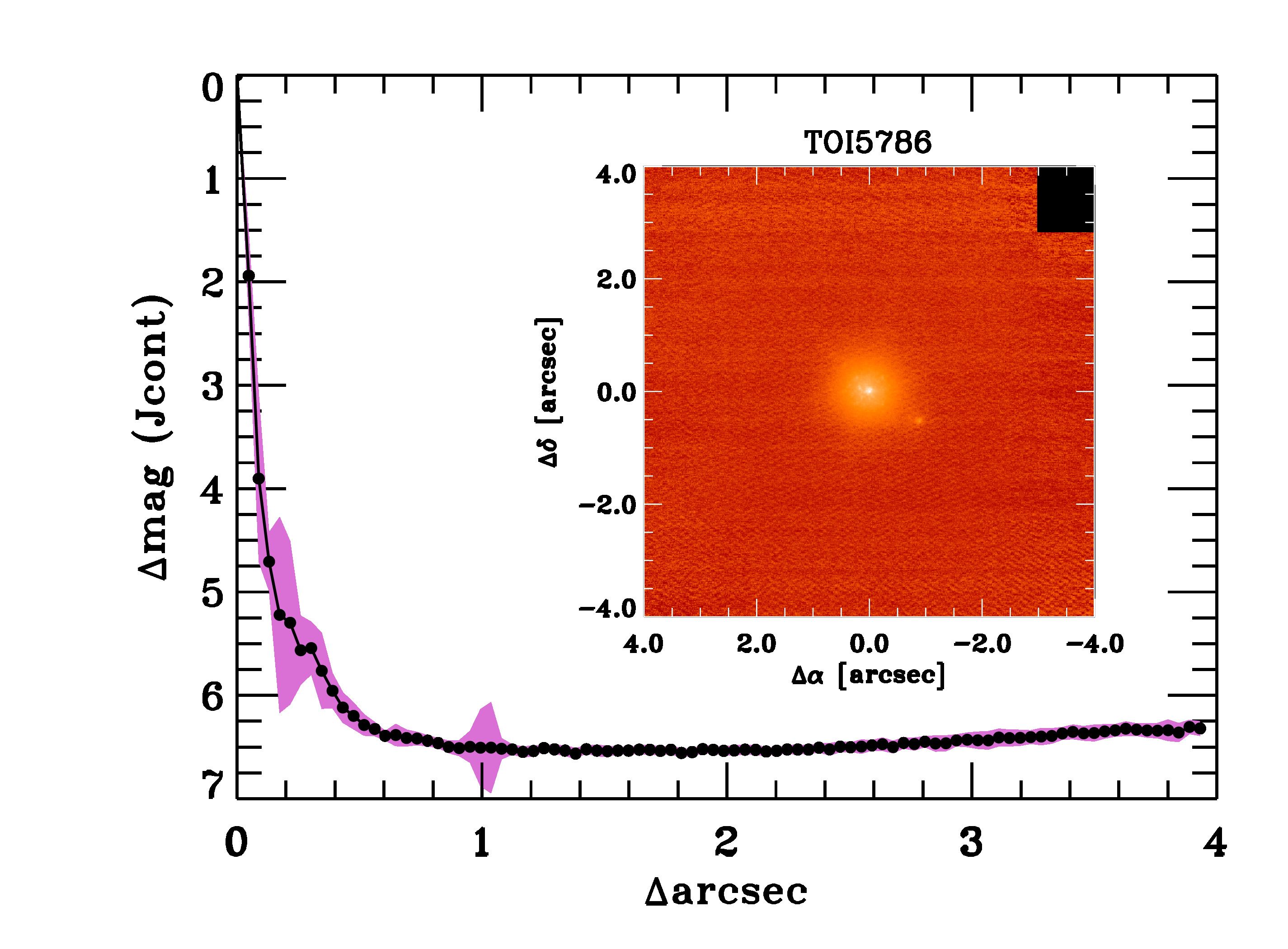} 
    \includegraphics[width=0.49\textwidth]{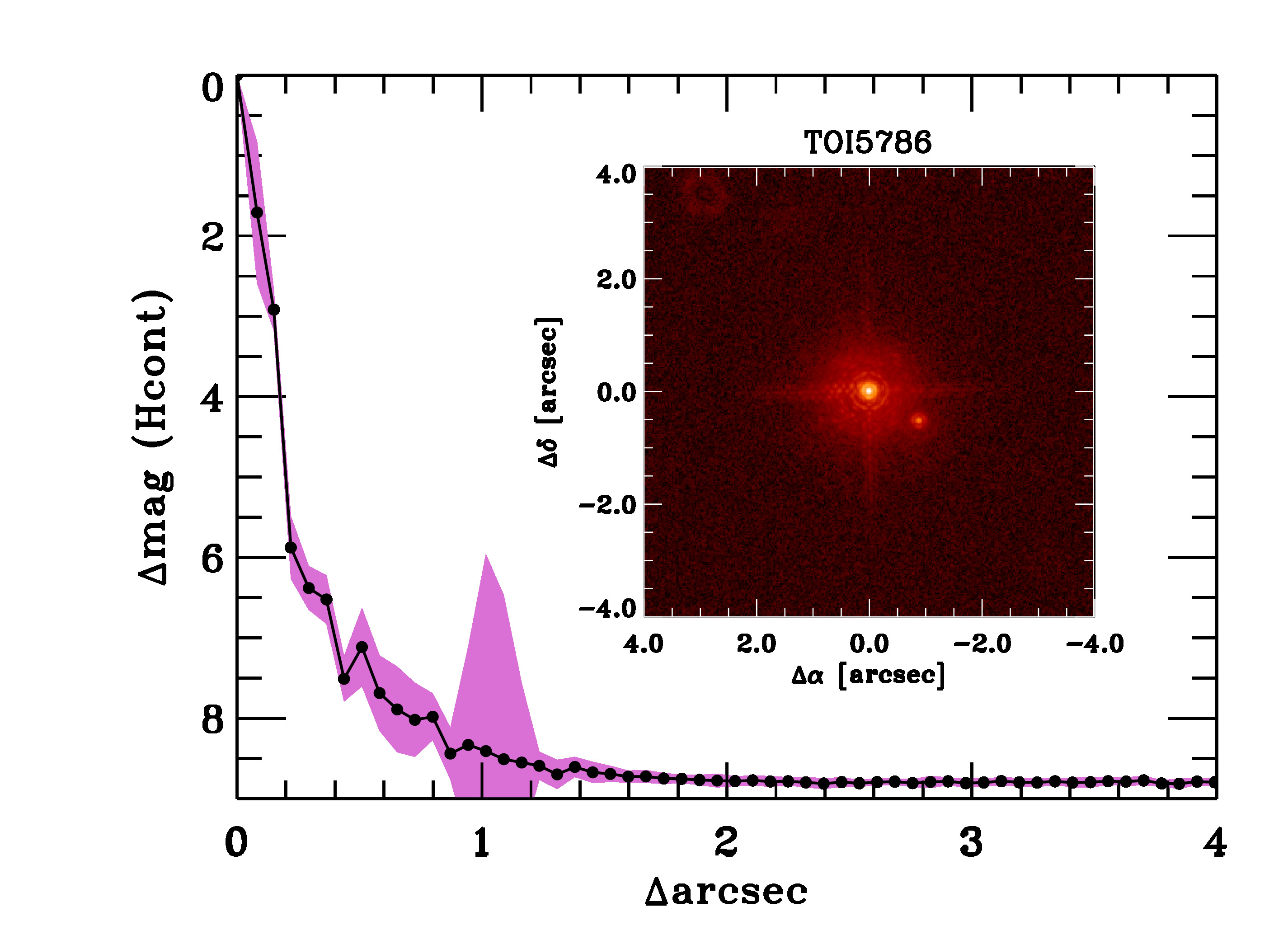} 
    \includegraphics[width=0.49\textwidth]{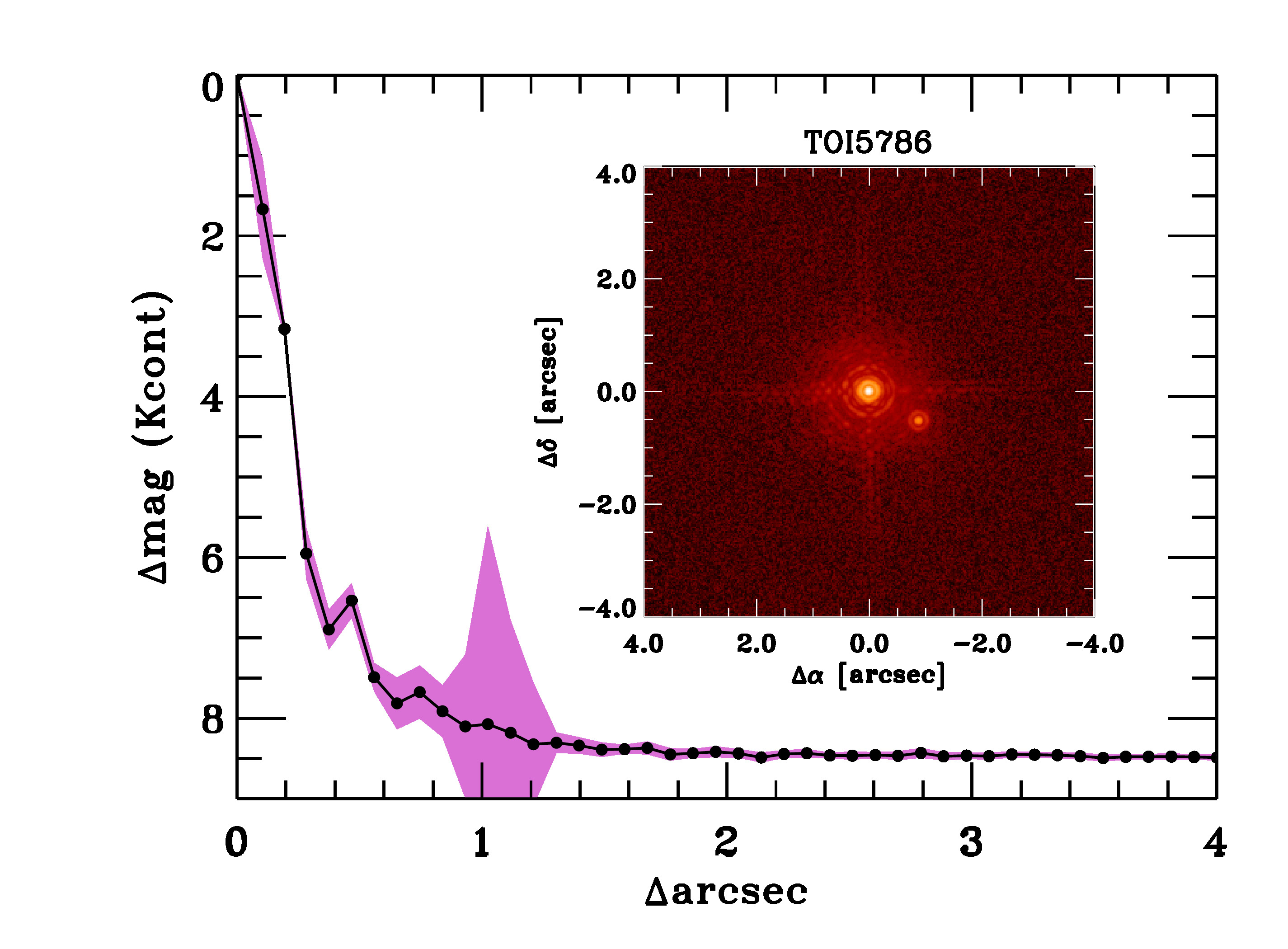} 
    \captionof{figure}{High-angular resolution imaging observations for TOI-5108 (upper left) and TOI-5786 (upper right, middle and lower plots). For TOI-5108 no secondary sources were detected. For TOI-5786 a secondary source at the separation $\sim 1\arcsec$ was detected.}
    \label{fig:dirim}
\end{minipage}
\section{SED Plots}
\begin{minipage}{\textwidth}
    \centering
    \includegraphics[width=0.45\textwidth]{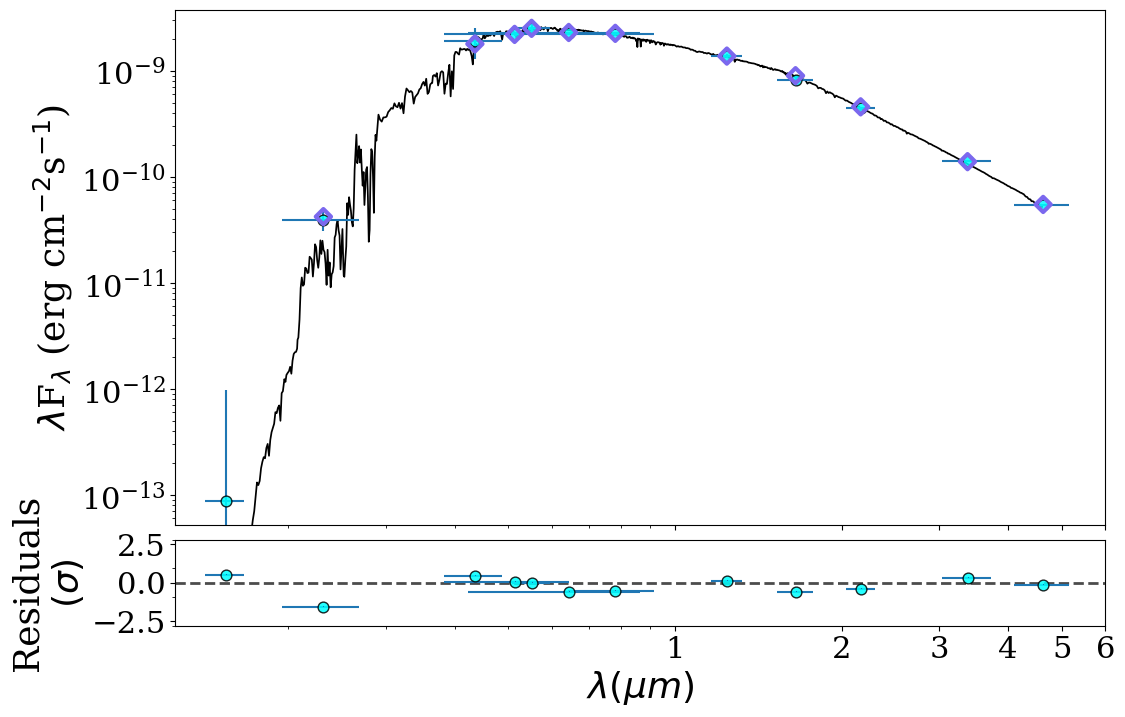}
    \includegraphics[width=0.45\textwidth]{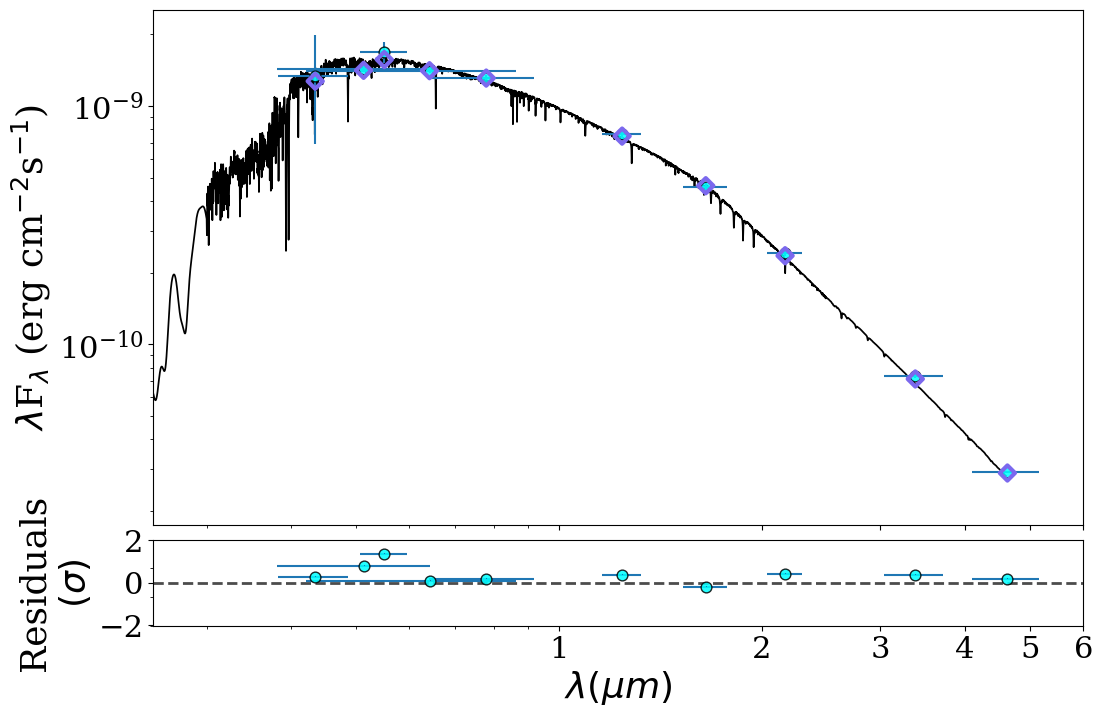}  
    \captionof{figure}{SED fits for TOI-5108 (left) and TOI-5786 (right). Blue points represent the flux values from available broadband photometry and the purple diamonds indicate the flux value of the synthetic photometry in the same passband. The best-fitting model is plotted in black. Residuals normalized to the errors of the photometry are shown below.}
    \label{fig:sed}
\end{minipage}
\end{appendix}

\end{document}